\documentclass[3p]{elsarticle}

\biboptions{longnamesfirst,sort&compress}
 
\usepackage{graphicx}
\usepackage{amssymb,amsmath}
 \usepackage{bm}

\newcommand{\mvisc}{\ensuremath{{  \mu}}}                               
                           
\newcommand{\visc}{\ensuremath{{\mvisc}}}

\newcommand{\ndLength}{\ensuremath{{  \lambda}}}

\newcommand{\meanu}{\ensuremath{{  \langle u_x \rangle }}}            
\newcommand{\grad}{ \ensuremath{\vec \nabla} }
\newcommand{\norm}[1]{ \ensuremath{|\!| #1} |\!| } 
\newcommand{\vecuprime}{\ensuremath{\delta \vec u}}
\newcommand{\uprime}{\ensuremath{\delta  u}}
\newcommand{\pprime}{\ensuremath{\delta p}}

\newcommand{\Occ}{\ensuremath{\mathcal R} }  
\newcommand{\uc}{u_c}            
\newcommand{\gc}{\ensuremath{{   g_c}}}           
\newcommand{\gref}{\ensuremath{{   g_0}}}            
\newcommand{\uref}{\ensuremath{{   u_0}}}

\newcommand{\au}[1]{\ensuremath{ {  \vec \omega_{#1} }}}

\begin{document}
  
\title{On the statistical properties of fluid flows with transitional power-law rheology in heterogeneous porous media
}

\author{Laurent Talon}
\address{Universit\'e Paris-Saclay, CNRS, FAST, 91405, Orsay, France.}

\begin{abstract}
In this work, we study non-Newtonian fluid flow in heterogeneous porous media.
We are interested in fluids presenting a specific change in rheology: Newtonian below a certain shear rate and power law above.
Since porous media generally exhibit strong spatial heterogeneity at large geological scales, we study the interaction between such inhomogeneity and the nonlinear rheology of the fluid.
The coupling between permeability heterogeneity and nonlinear rheology  significantly affects the flow. 
We are particularly  in the statistical properties of the velocity field (mean, variance, correlation, etc).

Depending on the imposed mean pressure gradient, three macroscopic flow regimes are identified.
For a low or high average pressure gradient, the average flow rate increases linearly or according to a power law, respectively.
In the latter regime, we observe that the velocity field is more heterogeneous for shear-thinning fluids than for shear-thickening fluids.
This is corresponding to a channeling effect of shear-thinning fluids.

The intermediate regime corresponds to a progressive and inhomogeneous change of the local rheology.
This transient regime is then characterized in terms of pressure gradient range. The flow field is also analyzed statistically.
The spatial distribution of the regions above the rheology threshold shows interesting statistical properties.
For instance, they  exhibit multiscale characteristics (fractal), similar to other critical systems (percolation, avalanches, etc.).
If the distribution of their area follows a power-law, the exponent is independent of the disorder. This suggests a kind of "universality" in this problem. More surprisingly, even though some statistical properties are independent of the parameters, an interesting abrupt rotation of the correlations is found for a particular set of parameters. This is explained by using some symmetries of the problem.
\end{abstract}
 
\maketitle  
 
\section{Introduction}
Many natural or industrial fluids exhibit non-Newtonian behaviors, \citep{bird87a,barnes89,coussot05}, 
 they are thus found in many applications related to porous or fractured media.
 A very important application is certainly enhanced oil recovery (EOR) (see \cite{sorbie91}). The most standard way to recover oil is to inject water to force the oil to move (waterflooding). 
The efficiency of waterflooding is however related to the uniformity of the displacement front. The more uniform the front, the greater the amount of oil to be moved. In contrast, if the preferential paths are formed, they will bypass certain regions of the porous medium leaving the oil in place. Two main reasons can cause this bypass. The first one comes from the heterogeneity of the permeability field. The fluid indeed tends to flow into the highest permeability regions and avoid the lower permeability ones.
The second reason comes from the viscosity contrast. If the displacing fluid has a lower viscosity than the displaced one, the front can destabilize and form fingerings as described by  Saffman and Taylor \cite{saffman58} or Homsy \cite{homsy87}.
To limit this bypass effect, a solution used by the oil industry is to add a polymer (e.g. Xanthan) to the displacing water (see \cite{sorbie91}). Another advantage of improving the viscosity ratio lies in the fact that it also reduces the effects of heterogeneity in the permeability field.
The main difficulty of this process is that most of the injected polymers have a non-Newtonian rheology. 
This is the case for example of Xanthan which is Newtonian at low shear rates  but is  shear thinning at higher shear rates. 
However, as has been shown in pore networks \cite{shah95} or in fractures \cite{bessonov16}, nonlinear rheology can amplify or dampen heterogeneities.
The objective of this paper is to further investigate the interplay between the macroscopic inhomogeneity of a permeability field and the flow of a nonlinear fluid. We focus particularly on the change in behavior of the viscosity.

Another interesting application of non-Newtonian fluids is the understanding of blood flow in capillary networks. Indeed, blood, which must be considered as a suspension, presents a non-Newtonian viscosity: shear thinning or yield stress. \citep{boyd07, bessonov16}.
Non-Newtonian fluid is also present in the proppants used for hydraulic fracturing or the mud produced by drilling wells \cite{bittleston02,frigaard17}. They are also commonly used for fracture sealing (\emph{e.g} cements, polymers, etc.) \citep{tongwa13}.

A very recurrent problem when dealing with porous media is that of upscaling.
If the equations of motion are generally well known at the pore scale (typically $\sim 10^{-3} \; \rm{m}$), a particular interest is to understand the flow at much larger scales ($\sim 1 - 10^3\; \rm{m}$)
This is usually done by deriving constitutive equations for average quantities at an intermediate scale and is illustrated by the famous Darcy's law for Newtonian fluids, which relates linearly the  mean flow rate to the macroscopic gradient of pressure.

At the microscopic level, Newtonian fluids obey the Stokes equation (neglecting the inertia) :
\begin{equation} 
 \vec 0  = - \grad p + \mu \Delta \vec v,
\end{equation}
where $\vec v$ is the fluid velocity, $p$ is the pressure and $\mu$  the viscosity.
Averaging the velocity and pressure field over a large number of pores results in Darcy's law\citep{darcy56}:
\begin{equation}
  \vec u  = - \frac{\kappa}{\mu} \grad P,
\end{equation}
where $\vec u$ is the mean velocity, $\grad P$ an averaged pressure gradient and $\kappa$ the permeability of the porous medium which depends on its structure.

At the geological scale, the type of rock may however spatially vary leading to a macroscopic heterogeneous permeability field. The large-scale flow obeys then to the heterogeneous Darcy's law:
\begin{equation}
 \vec u = - \frac{\kappa(\vec r)}{\mu} \grad P \; \; \; {\rm and}   \; \; \; \vec \nabla . \vec{ u} = 0.
\end{equation}

It is also worth recalling that a very similar equation is used for solving flow in  rough fractures, usually referred as the Reynolds equation \citep{reynolds86,zimmerman91a,mourzenko95}. 
Indeed, in a fracture with varying opening and under the lubrication approximation (small thickness and small variation of the opening), the flow obeys:
\begin{equation}
 \vec q(x,y) = - \frac{\kappa(x,y)}{\mu} \grad_{2D} P \; \; {\rm et} \; \vec \nabla_{2D} . \vec {q} = 0,
\end{equation}
where the fracture is in the $(x,y)$-plane, $b(x,y)$ represents the local opening,
$\kappa(x,y) = \frac{b^3(x,y)}{12}$ and $\vec q(x,y)= \int_0^{b(x,y)} \vec u \; dz$ is the flow rate.
The fracture can thus be treated as a 2D porous medium. There is, however, a small caveat because the dimensions are a slightly different. In a porous medium, Darcy's law implies an average velocity ($m.s^{-1}$) and the permeability has the dimension of $m^2$, whereas in a fracture, $\vec q$ is a volumetric flux per unit length ($m^2.s^{-1}$) and the ``permeability'' has the dimension of $m^3$.

The influence of the heterogeneity of a permeability field (or fracture) has also been the subject of a considerable amount of work, starting with the work of Matheron \cite{matheron67}, Gelhar and Axness \cite{gelhar83} or Dagan \cite{dagan84}. One can also mention the review by Renard and de Marsily \cite{renard97}.
Since the fluid prefers to flow in high permeability regions and avoid low permeability ones, the heterogeneity strongly influences the velocity field.
The central question is to understand how the permeability distribution affects the velocity field, in particular its average rate and the magnitude of its heterogeneity. The latter is specifically important for describing the transport of species (e.g., pollutants) in the subsurface \cite{dagan82}.

All studies mentioned above  apply to Newtonian fluids. Therefore, the following question naturally arises: how should these approaches be modified when considering non-Newtonian fluids?
Although there exists a very large variety of non-Newtonian fluids \citep{bird76,bird87a, coussot05},
several similar approaches are commonly used in the case where it exists a simple relationship between shear rate $\dot \gamma = \sqrt{ \frac{1}{2}  {\rm tr} \;  [ ( \grad \vec v + \grad \vec v^{T}) ^2 ]} $ and  viscosity $\mu(\dot \gamma)$, or equivalently between $\dot \gamma$ and shear stress $\tau(\dot \gamma)$, with $ \tau(\dot \gamma) = \mu(\dot \gamma) \dot \gamma$. 
One approach (see for instance \citep{christopher65,sadowski65, slattery67,hirasaki74, chauveteau82}) consists in determining an effective shear rate $\dot \gamma_{pm}$ to derive an effective viscosity.
Other ones are to establish an effective stress \citep{mckinley66}, or an average viscosity \citep{eberhard19}.
The common feature of these approaches is that they are based on the determination of mean effective  quantities.  
They can be synthesized using scaling arguments. Indeed, by defining a typical length scale $\tilde \lambda$ (pore size, grain diameter, $\sqrt{\kappa}$, etc.) and using the average flow rate $u$, a typical shear rate $\dot \gamma_{eff} \propto u / \tilde \lambda$ can be defined. A typical shear stress $\tau_{eff} \propto  \tilde \lambda \nabla P $ is also deduced from the pressure gradient. 
These quantities can then be used in the rheological function $\dot \gamma =f( \tau)$ to derive a generalization of Darcy's law in the form : 
\begin{equation}
    u \propto f( \nabla P ),
\end{equation}
where the pre-factors must be determined (experimentally, numerically or theoretically).
It is therefore expected that the flow/pressure curve will keep the overall shape of the rheological curve (see Fig. \ref{fig:sketch}). 
 Moreover,   Shah and Yortsos \cite{shah95} and Auriault \cite{auriault02} proposed a theoretical approach to homogenize the flow for  power law fluids in a periodic porous media. In both cases, the derived Darcy's law is also a power-law which is in agreement with  the  effective quantities approach.
   
\begin{figure}
\begin{center}
\includegraphics[width=0.49\hsize]{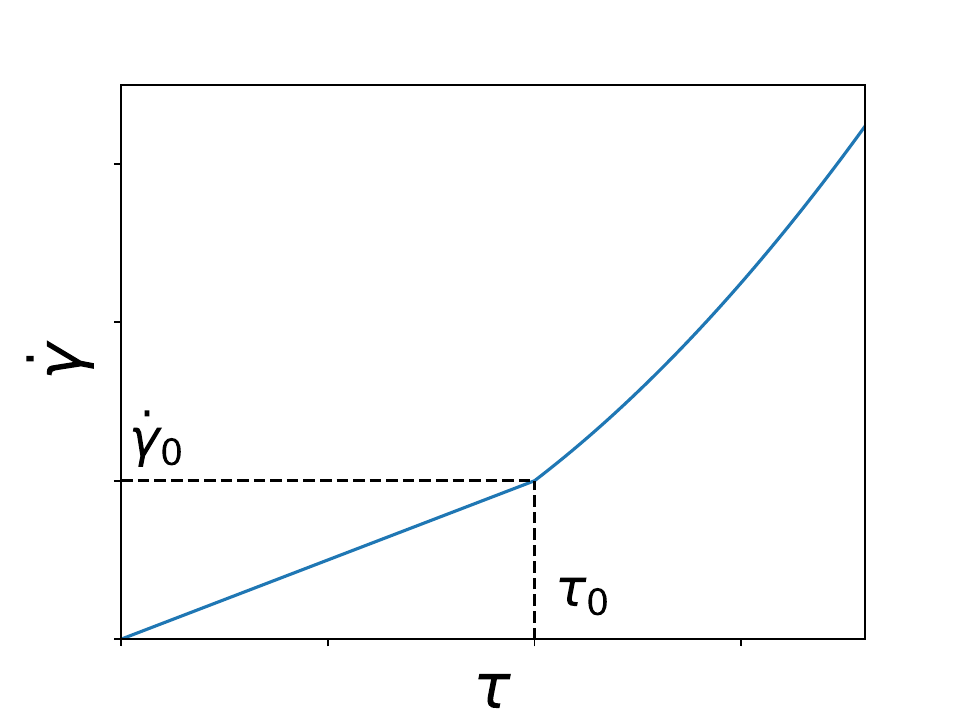}
\includegraphics[width=0.49\hsize]{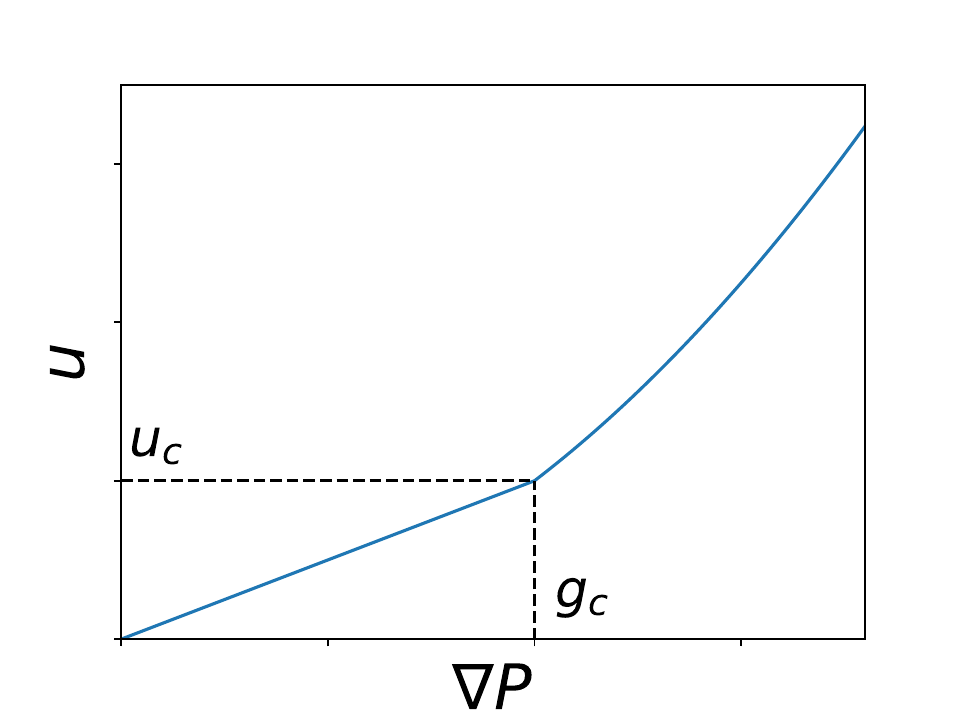}
\caption{\label{fig:sketch} Left: Relationship between shear rate and shear stress for a truncated rheology. For a shear rate lower than $\gamma_0$ (or a stress lower than $\tau_0$), the fluid is Newtonian. Beyond this thresholds the rheology is a power law.  Right: General Darcy's law, mean flow rate $u$ as function of the average gradient of pressure $\nabla P$ in porous media for the truncated rheology using the ``mean field'' rheology approach.}
\end{center}
\end{figure} 

The objective of this paper is to study rheologies that exhibit a change in behavior.
Indeed, many non-Newtonian fluids, such as Xanthan,  exhibit non-linear behavior only at high shear rate (or stress).
At low shear rates, they still behave like a Newtonian fluid.
To describe this rheology, a simple model is chosen which is the "truncated rheology", where the transition is sharp, at a given value of shear rate (or shear stress):
\begin{equation}
\label{eq:Carreau_Darcy_vec}
\left\{
\begin{array}{lll}
\tau = \mu \dot \gamma \; \; \; {\rm if}  \; \; \;& \dot \gamma < \dot \gamma_0 \\
\tau \propto \dot \gamma^n  \; \; \; {\rm if}  \; \; \;& \dot \gamma > \dot \gamma_0 \\
\end{array}
\right.,
\end{equation}
where $n$ is the flow index.

Using the ``mean field'' approach, as studied numerically by \cite{lopez03,zami-pierre16} for example, Darcy's law can be written as follows:
\begin{equation}
\label{eq:Carreau_Darcy_vec}
\left\{
\begin{array}{lll}
|\nabla P| =  \frac{\visc}{\kappa}  u & \; \; \; {\rm if}  \; \; \;&  u < \uc \\[\bigskipamount]
|\nabla P|  = \frac{\visc}{\kappa} ( \frac{u }{\uc} )^{n-1} u & \; \; \; {\rm if}  \; \; \;&  u > \uc, \\
\end{array}
\right.
\end{equation}
where $\kappa$ is the  permeability and $\uc$ a velocity threshold that depends on $\dot \gamma_0$ and the porous structure. The prefactor in the second equation is determined by continuity.
$\kappa$ and $\uc$ are thus parameters depending on the topology of the porous medium (pore size distribution, porosity, etc.).
At macroscopic geological scales, however, the structure of porous media is expected to vary, resulting in a heterogeneous permeability and $\uc(\vec r)$ field.
The main objective of this paper is therefore to study the influence of this field inhomogeneity on the flow for a fluid presenting a transient rheology.
Another application  could be also to evaluate the use of non-Newtonian fluids to characterize the degree of heterogeneity of a field. Indeed, as each location changes its viscosity behavior at a different flow rate, recording the evolution of the average velocity could potentially give indications on the permeability heterogeneity.

This article is structured as follows. Section \ref{sec:problem} is devoted to the presentation of the problem.
Section \ref{sec:numerics} contains the numerical results, where the different flow regimes are analyzed.
In particular, the transient regime presents interesting statistical properties.
Section \ref{sec:conclusion} is dedicated to discussions and conclusions.
Different appendices provide some mathematical properties of the nonlinear Darcy's law (perturbation expansion, symmetry) and also the numerical method used.

\section{Problem description - Governing equations \label{sec:problem}}

\begin{figure}
\begin{center}
        {\includegraphics[width=0.5\hsize]{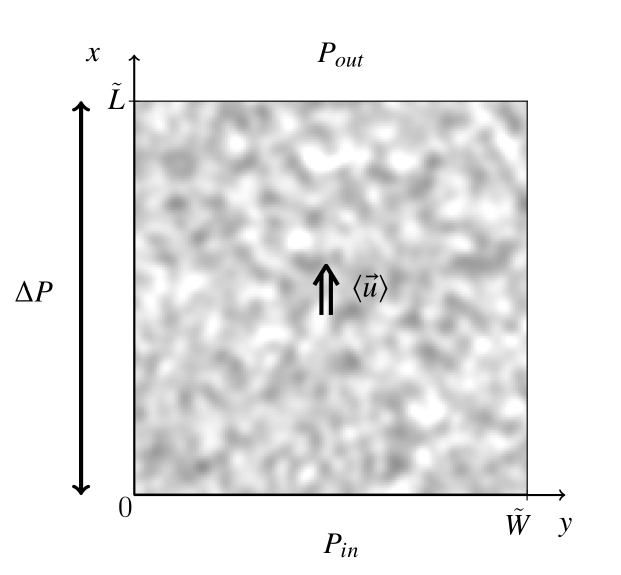}}
 \caption{\label{fig:sketch}Schematic of the studied system. The heterogenous domain has a size $\tilde L \times \tilde W$. The flow is driven by imposing a pressure difference between $x=0$ and $x=\tilde L$.}
  \end{center}
\end{figure}

To solve the flow field, a vector formulation of the non-Newtonian Darcy's law is required. Assuming that the medium is locally isotropic so that the mean flow is collinear and opposite to the mean pressure gradient, it follows:

\begin{equation}
\label{eq:Carreau_Darcy_vec}
\left\{
\begin{array}{lll}
\grad P = - \frac{\visc}{\kappa(\vec r)} \vec u & \; \; \; {\rm if}  \; \; \;& \norm{\vec u} < \uc(\vec r) \\
\grad P  = - \frac{\visc}{\kappa(\vec r)} \left[ \frac{\norm{\vec{u}} }{\uc(\vec r)} \right]^{n-1} \vec{u} & \; \; \; {\rm if}  \; \; \;& \norm{\vec u} > \uc(\vec r) \\
\end{array}
\right..
\end{equation}
In addition, the velocity field must satisfy mass conservation:
\begin{equation}
    \grad.\vec u =0
\end{equation}

The flow field is solved using a second order finite difference method combined with an augmented Lagrangian approach described in the appendix. In practice, the flow rate is determined by imposing a pressure difference $\Delta P$  between the inlet and the outlet (see Fig. \ref{fig:sketch}). However, it is more convenient to use the average pressure gradient $\langle \nabla P \rangle = \frac{ \Delta P}{\tilde L}$ as a control parameter, where $\tilde L$ is the length of the system and the mean operator is defined as $\langle.\rangle = \frac{1}{\tilde W \tilde L} \int .\;dxdy$.  The lateral boundary conditions are assumed to be periodic. By construction, $\langle \nabla P \rangle$ has therefore a constant direction, aligned with the x-axis.

The permeability field was chosen to be distributed according to a log-normal distribution which has been a common model since the work of Gelhar and Axness \cite{gelhar83}. It has the advantage of being consistent with the field data but also of allowing the mean and variance to be varied independently.
 The permeability field is obtained by generating a Gaussian field $\delta f$ of zero mean and given standard deviation $\sigma_f$. $\delta f$ has a correlation length $\ndLength$ (see \citep{kostenko19} for more details).
The permeability field is then given  by:
\begin{equation}
    \kappa = \exp{(f_0 + \delta f)}=\kappa_0 \exp{(\delta f)},
\end{equation}
where $\kappa_0 = \exp{(f_0)}$, is a parameter characterizing the average permeability of the medium.

The threshold field $\uc$ is expected to depend on both $\dot \gamma_0$ and the pore structure. $\dot \gamma_0$ is a characteristic of the fluid rheology and is therefore constant.
The pore structure may however vary spatially and  is  related to the permeability $\kappa$.
A simple relationship can be established using phenomenological arguments. For porous media with a typical pore size $d$, a scaling analysis leads to $\kappa \sim d^2$ et $\uc \sim \gamma_0 d$, which gives: $$\uc \sim \kappa^{1/2}.$$ 
This scaling has for example been evaluated in \cite{zami-pierre16}.
It is important to stress that this argument is very crude.
While this scaling law is certainly valid in the case of homothetic transformations, it is not necessarily applicable to more complex structural changes. In other words, this scaling law is most likely valid when changing the diameter of a packet of monodisperse beads (or sand). But it is probably more complicated if the nature of the medium changes radically, from a sandpile to a porous rock for example.

As mentioned in the introduction, two-dimensional Darcy's law can also be used to solve the flow in heterogeneous fractures. The scaling is then slightly different.
By defining $h$ as the opening, the local permeability leads to $\kappa \sim h^3$ and $\uc \sim \gamma_0 h^2$. It follows: $$\uc \sim \kappa^{2/3}.$$
 
The field $\uc$ is thus determined from the permeability field $\kappa(x,y)$ by assuming a more generic scaling law of the form:
\begin{equation}
\uc = A \kappa^{\gamma},
\end{equation}
where $A$ is a prefactor and $\gamma$ a parameter with $\gamma \in [0;1]$.

Although there are many different parameters, in this study we will focus on the disorder amplitude $\sigma_f$, the flow index $n$ and the exponent $\gamma$.
As discussed in \ref{sec:numerical_method}, the numerical method is particularly efficient for specific values of $n \in [1/3; 1/2; 2/3; 1; 3/2; 2; 3]$, which are the values used in this work.
The parameters $A$, $\mu$ and $\kappa_0$ will be kept constant with $A=1$, $\mu=0.1$ and $\kappa_0=1$.

Based on these parameters, a characteristic velocity $\uref = A \kappa_0 ^\gamma$ and a characteristic pressure gradient $\gref = \frac{\mu \uref}{\kappa_0}$ are defined and will be used for non-dimensionalization.
Basically, $\uref$ and $\gref$ represent the velocity and pressure gradient at which the system would change behavior if the field were homogeneous (\emph{i.e} $\sigma_f=0$).

\section{Numerical results \label{sec:numerics}}

\subsection{Simple case: Heterogeneous permeability - homogeneous critical velocities}

\begin{figure}
\begin{center}

        \includegraphics[width=0.8\hsize]{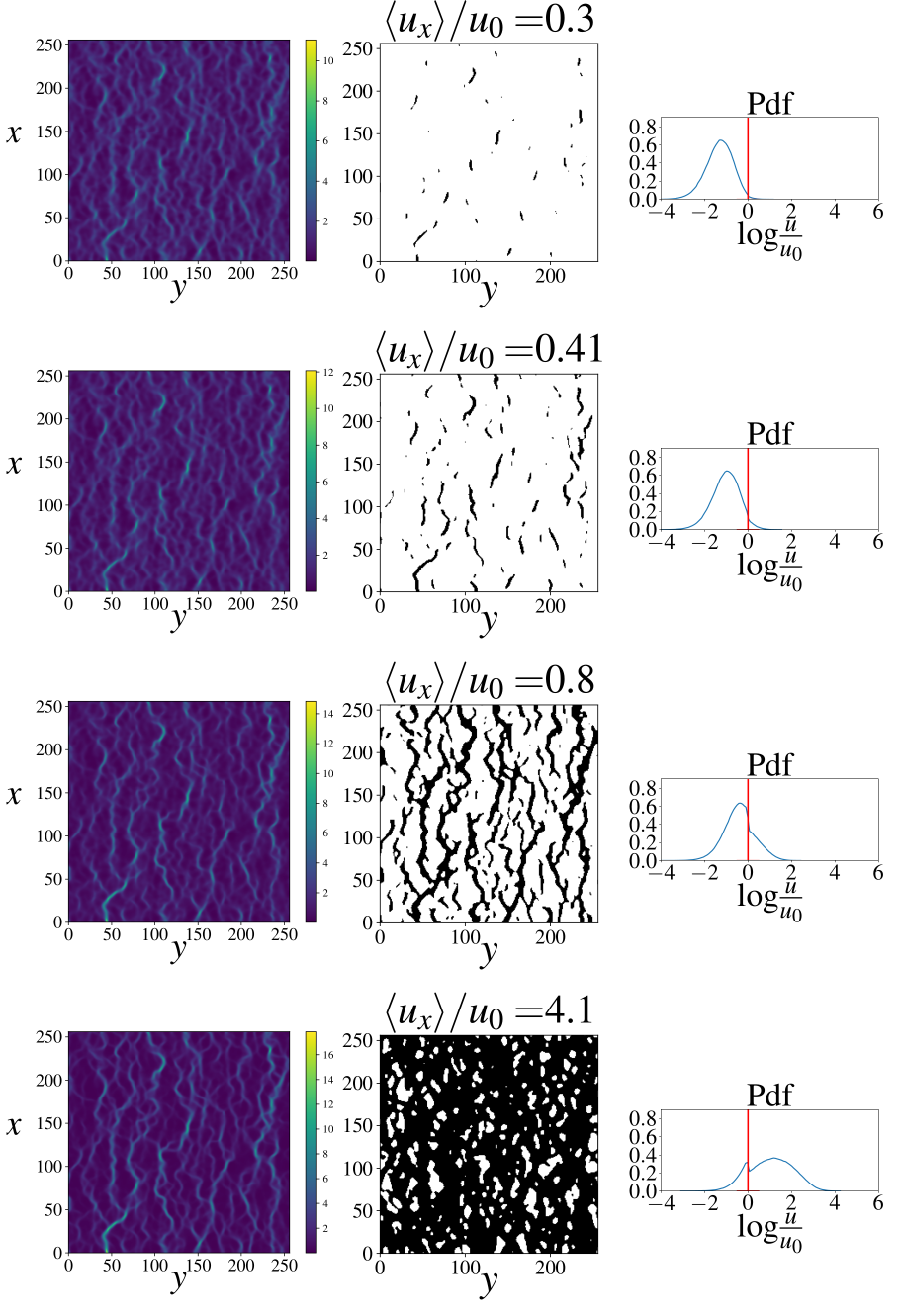}

	\caption{\label{fig:flow_example_homuc}  
	For a shear-thinning  truncated rheology $n=1/2$ in a heterogeneous permeability field, with $\sigma_f =1$ with a uniform velocity threshold $u_c$ (\emph{i.e} $\gamma=0$).
    Left column: Velocity field colormap  for different imposed mean pressure gradients (increasing from top to bottom). The average flow direction is from  top to bottom.
	Middle column: Corresponding regions (in black) above the threshold.
	Right column: Corresponding velocity density distribution function (blue) along with the threshold velocity $u_c$ (red vertical line).
	}  
	\end{center} 
\end{figure} 
  
\begin{figure}
\begin{center} 
        \includegraphics[width=0.8\hsize]{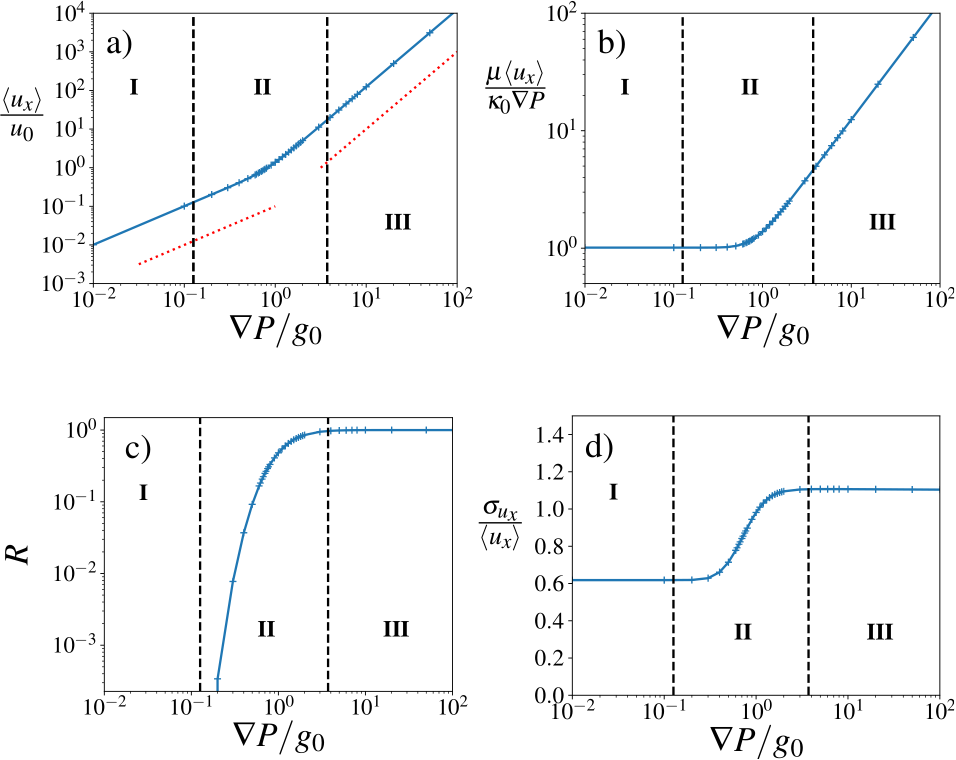}
	\caption{\label{fig:curve_homk} 
	For a shear-thinning  truncated rheology $n=1/2$ in a medium with a heterogeneous permeability field with $\sigma =1$ with a homogeneous threshold field ($\gamma=0$).
	As function of the mean pressure gradient:
	{\bf a}: mean velocity $\langle u_x \rangle$.
	The red dashed lines represent the power law $\langle u_x \rangle \propto \langle \nabla P \rangle$ and $\langle u_x \rangle \propto \langle \nabla P \rangle^{1/n}$.
	{\bf b}: mean velocity rescaled with the Newtonian Darcy's law.
	{\bf c}: percentage of regions above the threshold ($\norm{\vec u}>\uc$).
	{\bf d}: relative standart deviation of the velocity field $\frac{\sigma_u}{\meanu}$.
	Vertical dashed lines represent the transitions of the flow regimes (I, II and III).
	}
	\end{center}
\end{figure} 
 
In order to describe the problem qualitatively, a simplified version is presented here, where the permeability is heterogeneous but the critical velocity field $\uc$ is homogeneous (\emph{i.e} $\gamma=0$). 

Figure \ref{fig:flow_example_homuc} (left) shows the flow field for different average pressure gradients $\langle \nabla P \rangle$.
For a very small pressure gradient, the flow field is heterogeneous but all velocities are below the (single-valued) threshold $\uc=\uref$ (figure not shown).
As the pressure gradient increases, the velocity field increases as a whole. At some point, some regions reach the threshold (Fig. \ref{fig:flow_example_homuc}, top).
At these locations, the viscosity is changed, which disturbs the surrounding flow field. This modification  therefore favor or disfavor the velocity in the close vicinity. 
Here, the fluid is shear-thinning $n<1$, so the viscosity is locally decreased.
The flow is then increased downstream and upstream, while it is decreased on the lateral sides. 
Thus, some correlations of regions that have changed flow regime can be expected. For shear-thinning (resp. thickening), the regime change should thus be correlated along the flow direction (resp. perpendicular to it).
As the pressure gradient increases, more and more regions change their behavior until the entire domain is in the non-Newtonian regime.
We can also note how the velocity distribution is crucially altered as it passes through the threshold value in Fig. \ref{fig:flow_example_homuc} (right column).

Fig. \ref{fig:curve_homk} shows the evolution of various interesting quantities.
Fig. \ref{fig:curve_homk}.a displays the average velocity as a function of the average pressure gradient.
As expected, for low pressure, the average flow rate varies linearly with the average pressure gradient. Above a certain value, the curve starts to slowly deviate from the linear trend to a power law type trend.
Three flow regimes are thus identified. For the first one (regime I), the average flow rate increases linearly.
At a high flow rate (regime III), the mean flow follows a power law with exponent $\meanu \propto \langle \nabla P \rangle^{1/n} $.
Between these two regimes, a transitional one (regime II) is observed.
 This change in behavior can be magnified by plotting the ratio $\meanu/\langle \nabla P \rangle$ for example (Fig.\ref{fig:curve_homk}.b).
Another convenient quantity to characterize this transition is the percentage of regions in the nonlinear regime $\Occ(\langle \nabla P \rangle) \in [0,1]$ (Fig. \ref{fig:curve_homk}.c).
A final very important quantity, shown in Fig. \ref{fig:curve_homk}.d, is the relative standard deviation of the velocity field, $\frac{\sigma_u}{\meanu}$, which characterizes the heterogeneities of the flow and is of great importance, for example, in the problem of species transport.
Here we observe that regime I and III correspond to a plateau value of the standard deviation.

This example shows how permeability heterogeneity affect the flow regime.
Since regions with high permeability have higher local velocity, they are more likely to reach the nonlinear viscosity regime. However, the more general case $\gamma \neq 0$ is more complex because the velocity threshold is also distributed in space. Additionally, regions of higher permeability also correspond to a higher velocity threshold. A competition between these two effects is thus expected.

\subsection{Influence of the amplitude of the field heterogeneities}

\begin{figure} 
\begin{center}
        \includegraphics[width=0.8\hsize]{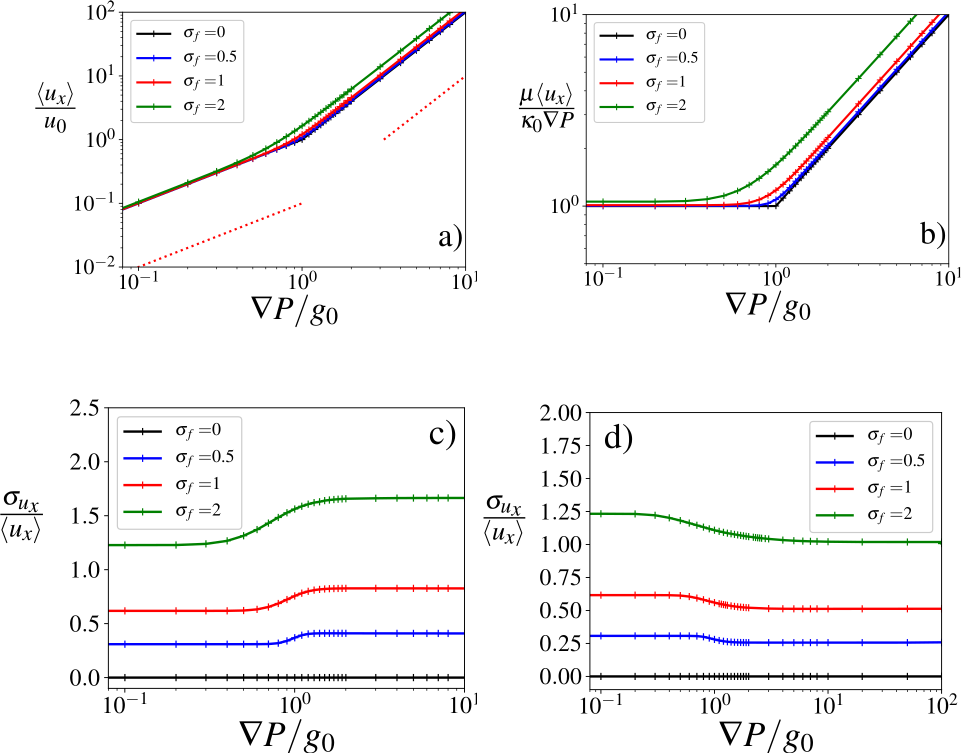}
 \caption{
 \label{fig:flowrate_example} For a truncated shear thinning fluids with $n=1/2$ and $\gamma = 1/2$:
 {\bf a}: non-dimensional mean flow rate. {\bf b}: mean flow rate normalized by the Newtonian Darcy's regime. {\bf c}:  relative standart deviation of the velocity field as function of the non-dimensional mean gradient of pressure for different amplitude of heterogeneity $\sigma_f$.
 {\bf b}: Relative standart deviation of the velocity field as function of $\nabla P$ for a shear-thickening fluid ($n=2$).
 }  
 \end{center} 
\end{figure}

Fig. \ref{fig:flowrate_example}.a displays the average flow rate as a function of the average pressure gradient $\langle \nabla P \rangle$ for a shear thinning fluid, $n=1/2$, for $\gamma=1/2$ and for different magnitudes of heterogeneity $\sigma_f$.
Similarly to the homogeneous velocity threshold case, a transient behavior is observed around $\langle u_x \rangle \sim \uref$ and $\langle \nabla P \rangle \sim \gref$. 
In Regime I, the flow rate follows a linear behavior $\meanu \propto \langle \nabla P \rangle$, while in regime III, it follows a power law $\meanu \propto \langle \nabla P \rangle^{1/n}$.
The disorder then smoothes the transition while increasing its range. For $\sigma_f=0$, the transition is abrupt, while for $\sigma_f$ high, the transition is smoother and  extending over half a decade.
This effect is more apparent after normalizing the mean flow by Darcy's Newtonian law $\frac{\mu \meanu}{\kappa_0 \nabla P}$ (Fig. \ref{fig:flowrate_example}.b).

The second effect of heterogeneity is to significantly increase the flow rate in the high pressure gradient regime, where the flow rate is almost doubled between $\sigma_f=0$ and $\sigma_f=2$. 
This effect will be further detailed later  as it can be predicted using a perturbation expansion approach (\ref{sec:expansion}). 

Fig. \ref{fig:flowrate_example}.c plots the relative standard deviation of the  flow field, $\frac{\sigma_{ux}}{\meanu}$, as a function of the mean pressure gradient and for different $\sigma_f$.
Like previously, we observe a change of plateau when changing the flow regime. And the values of these plateaus increase with the magnitude of the  permeability heterogeneity $\sigma_f$.
Here again, the range of the transition between these two asymptotes depends on the heterogeneities of the porous medium, abrupt at low $\sigma_f$ while smoother at high $\sigma_f$. We can also notice that these curves are very symmetrical. In fact, these curves can be fitted remarkably well with a hyperbolic tangent function (in semi-log representation).
 
 Fig. \ref{fig:flowrate_example}.d  shows the normalized standard deviation of the velocity as a function of the applied pressure gradient,  for a shear-thickening fluid ($n=2$). 
The most notable difference is that the heterogeneity of the velocity field decreases in regime III compared to regime I.
Shear-thickening fluids therefore attenuate the permeability field heterogeneity while shear-thickening fluids enhance it.  

\begin{figure}
\begin{center} 
        \includegraphics[width=0.8\hsize]{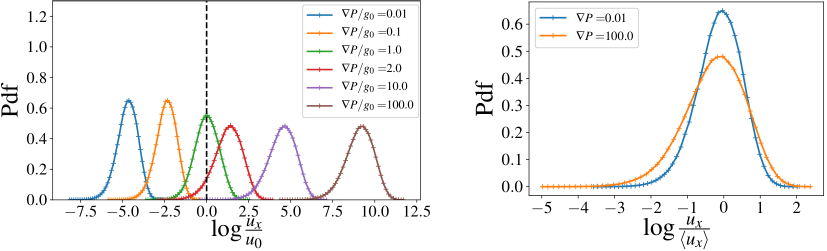}
 \caption{\label{fig:pdf_u_GP} For a shear-thinning fluid, $n=1/2$ and $\gamma=1/2$. Left: Evolution of the probability distribution function (PDF) of the velocity as function of the applied mean pressure gradient. The vertical dashed line represents the mean threshold  velocity $\uref$. Right:  normalized velocity distribution $\au{1}$ and $\au{n}$, obtained respectively  at  $\nabla P/\gref=10^{-2}$ (regime I) and $\nabla P/\gref=100$ (regime III). }
 \end{center}
\end{figure}

We have seen that the normalized standard deviation is constant in  regimes I and III.
It is noteworthy to mention that this is also the case for all  normalized moments of the velocity distribution (Skewness, Kurtosis, etc.).
In fact,  the velocity distribution actually keeps a constant shape in both regimes as observed in Fig. \ref{fig:pdf_u_GP}.
The distribution is only shifted when the mean pressure gradient varies.
As a result, the field $\vec u/\meanu$ is constant independently of the amplitude of the applied pressure gradient in both regimes.
We can therefore define a constant vector field : $\au{m}(x,y) =   \vec u^{PL,m}(x,y)/\langle u_x^{PL,m} \rangle,$ where $\vec u^{PL,m}$ is the velocity field of a power-law fluid with flow index $m$ ($m=1$ in regime I and $m=n$ in regime III).

\subsection{Statitical properties of regime I and III}

The flow field in regimes I and III is governed by the equation of  a power-law fluid.
For such fluid, it is however possible to determine the mean flux and standard deviation of the velocity using a perturbation expansion approach. 
The principle is to extend the work of \cite{gelhar83} on Newtonian fluids to power law fluids. It consists in expanding the permeability, pressure and velocity field around the mean value and in assuming that the deviation terms are small for sufficiently small $\sigma_f$. The complete calculation is provided in  \ref{sec:expansion} for a governing equation of the form :
$- \grad P = c(\vec r) \norm{\vec u}^{n-1} \vec u,$
 where we define $\vec r=(x,y)$ for the sake of conciseness.
In the case of the truncated rheology, we have $c(\vec r) = \frac{\visc}{\kappa(\vec r) \uc^{n-1}}$.
Because  $\kappa$ and $\uc$ are distributed according to a log-normal distribution, it is also the case for $c(\vec r)$ with: $\sigma_{\log c}  = ( 1 + \gamma (n-1)) \sigma_f$.
It follows:
\begin{equation}
\label{eq:siguth}
\frac{\sigma^2_{u_x}}{\langle u_x \rangle^2} = \left[ 1 + \gamma (n-1) \right]^2 \sigma_f^2  \frac{1 + 2 \sqrt{n}}{ 2 (1+ \sqrt{n})^2n^{3/2}  },
\end{equation}
and
\begin{equation}
\label{eq:mean_U_sigma_f}
 - \frac{\langle u_x \rangle }{D_0 \langle \nabla P \rangle^\alpha} = 1 +  \sigma_f^2 \left[ \alpha + \gamma (1-\alpha) \right]^2 \frac{\sqrt{\alpha} -1 }{2(\sqrt{\alpha}+ \alpha)}
\end{equation}
with $\alpha=1/n$ and $D_0 = \visc^{-\alpha} \kappa_0^{\alpha + \gamma (1 - \alpha)} A^{1-\alpha}$.

These results confirm qualitatively the previous observations that lower $n$  increase the mean flow rate and the flow heterogeneity.
As observed in pore network model \cite{shah95}, shear-thinning fluids are indeed more channelized and thus more heterogeneous.
The limit $n \rightarrow 0$ is interesting because it corresponds to a pure plastic flow (viscoplastic fluid without viscosity). The flow field is then expected to converge to a single flow path as for the Bingham fluid in the large Bingham number limit \cite{kostenko19}.
However, the expansion of the perturbations predicts a divergence of the standart deviation of the velocity at $n=0$.
Since the calculations assume small perturbations for each field, they are expected  to fail for a sufficiently small value of $n$.
On the other hand, increasing $n$ decreases the average velocity and the heterogeneity of the flow. The expansion approach should thus be better.

\begin{figure}
\begin{center}
        \includegraphics[width=0.8\hsize]{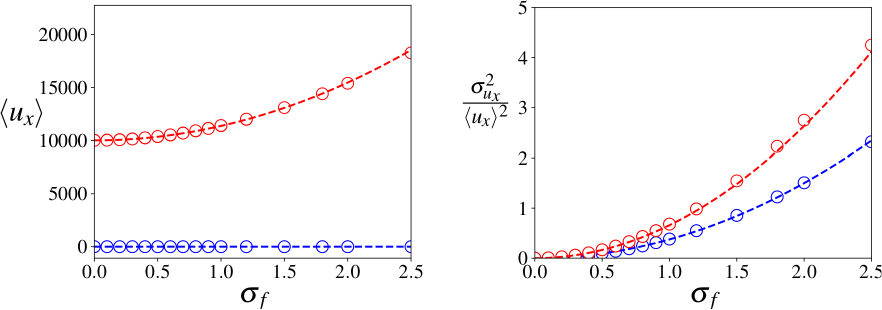}
 \caption{\label{fig:flow_rate_asymptotic} For a shear thinning fluid $n=1/2$ and $\gamma = 1/2$. Left:  mean velocity versus the amplitude of the heterogeneity $\sigma_f$ at low applied pressure drop $\nabla P/\gref=1e-2$ (blue) and high pessure drop $\nabla P/\gref=100$ (red). Right: relative standard deviation as function of $\sigma_f$. Circles represent the simulations and the dashed line is the perturbation expansion prediction eqs. (\ref{eq:siguth}) and (\ref{eq:mean_U_sigma_f}).}
 \end{center}
\end{figure}

Fig. \ref{fig:flow_rate_asymptotic} compares the numerical mean and standard deviation of the flow field for a sufficiently small
(resp. large) applied pressure gradient $\langle \nabla P \rangle/\gref=0.01$ (resp. $\langle \nabla P \rangle/\gref=100$) against the predictions of eqs. (\ref{eq:siguth}) and (\ref{eq:mean_U_sigma_f}).
Both figures show a very good agreement between the analytical predictions and the simulations, even for a heterogeneity amplitude as high as $\sigma_f \sim 2$, which is quite significant for a lognormal distribution.
An expansion of order $2$  seems thus sufficient to predict both the mean flux and standard deviation of the flow field in regimes I and III.

\subsection{Pressure range of regime II}

 \begin{figure}
  \begin{center}
          \includegraphics[width=0.8\hsize]{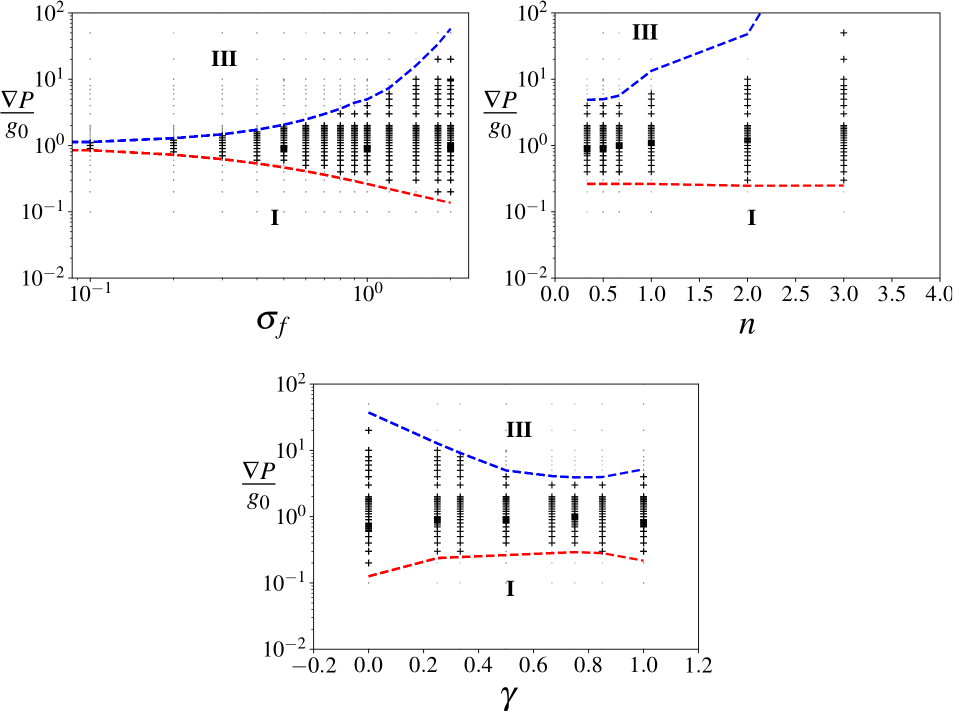}
  \caption{\label{fig:range_transition} 
Regime diagram of the system as function of different parameters. Crosses represent systems in regime II while the dots are the system in either regime I or III. The intermediate flow regime is defined as  $\Occ \in[10^{-5}, 1-10^{-5}]$.
  Top left: diagram for different applied pressure gradient and amplitude of heterogeneity $\sigma_f$, the other parameters are $(n,\gamma)=(1/2,1/2)$.
  Top right: diagram for different applied pressure gradient and flow index $n$, with $(n,\sigma)=(1/2,1)$. Bottom: diagram for different applied pressure gradient and  parameter $\gamma$.
  The red and blue dashed line represent respectively the bounds predicted by eqs. (\ref{eq:P1_bound}) and (\ref{eq:P2_bound}).}
    \end{center} 
\end{figure}

We now discuss the range of the transient regime II.
Fig. \ref{fig:range_transition} represents different phase diagrams of the system as a function of the mean pressure gradient while varying $\sigma_f$, $\gamma$ or $n$.
The crosses represent the system in the intermediate regime whereas the dots represent the system in regime I or III.

Fig. \ref{fig:range_transition}.a displays the evolution of the transient regime as a function of  $\sigma_f$.
This figure mainly confirms the previous observation that the range increases significantly with $\sigma_f$ because the velocity field is more heterogeneous.
Considering that the plot is on a logarithmic scale, the growth is in fact very significant as the range increases faster than a power-law.

Fig. \ref{fig:range_transition}.b shows the variation of the pressure range of regime II as a function of the rheological index $n$.
The transition range becomes narrower with increasing $n$.
This result is  contradictory to the previous observation that shear thickening fluids have lower velocity heterogeneity than shear thinning ones. 
It can also be noted that the lower pressure limit is not really affected by the value of $n$.

The influence of $\gamma$ is also not intuitive (Fig. \ref{fig:range_transition}.c) because the pressure range exhibits a non-monotonic behavior. The pressure range initially decreases with $\gamma$ but increases again above a certain value of $\gamma$.

The reason behind the last two observations is that the transition depends not only on the velocity distribution but also on the distribution of $u_c(\vec r)$.  The connection between both distributions is not obvious because regions of higher velocity are more likely to have higher permeability, and thus a higher threshold.
One way to highlight this phenomenon is to estimate the limits of regime II by exploiting the invariance of the velocity field in regimes I and III.

Starting at a very low flow rate, all regions are in the Newtonian regime and thus  $\norm{\vec u(\vec r)} = \meanu \norm{\au{1}(\vec r)}$.
When increasing the average flow rate, the first location that changes  its rheology occurs where $\norm{\vec u(\vec r)} = \uc(\vec r)$. It follows the condition of this first occurrence:
$ \max{\frac{\meanu \; \norm{\au{1}(\vec r)}}{\uc(\vec r)} } = 1.$
This means that the change of regime will start at the average velocity $\meanu = u_1$ satisfying:
\begin{equation}
 \frac{1}{u_1}= \max{\frac{ \norm{\au{1}(\vec r)}}{\uc(\vec r)} }.
\end{equation}

Similarly, starting with a high pressure gradient, all regions are in the power-law regime and  $\vec u(\vec r)/\meanu=\au{n}(\vec r)$  is  constant.
While decreasing $\meanu$, the first point changing its behavior occurs when $\meanu = u_2$, with:
\begin{equation}
 \frac{1}{u_2} = \min{\frac{\norm{\au{n}(\vec r)}}{\uc(\vec r)}}. 
\end{equation}

Assuming that at order zero $ \nabla P_1 \simeq \frac{\visc}{\kappa_0} u_1$ and $\nabla P_2 \simeq \visc {\kappa_0}^{-1 -\gamma (n-1)}A^{1-n} u_2^n $, it leads to an estimation for the pressure bounds of the transient regime:

\noindent
Lower bound:
\begin{equation}
 \nabla P_1 = \frac{\visc}{\kappa_0 \max{\frac{ \norm{\au{1}(\vec r)}}{\uc(\vec r)} }}
 \label{eq:P1_bound}
\end{equation}
Upper bound:
\begin{equation}
 \nabla P_2 = \frac{\visc {\kappa_0}^{-1 -\gamma (n-1)}A^{1-n}}{ \left(\min{\frac{ \norm{\au{n}(\vec r) }}{\uc(\vec r)}  }\right)^n}.
 \label{eq:P2_bound}
\end{equation}

\begin{figure}[ht]
 \begin{center}
         \includegraphics[width=0.4\hsize]{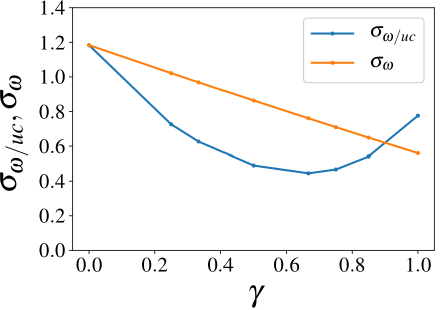}
  \caption{\label{fig:sigma_au} Standart deviation of the field $\frac{ \au{n}(\vec r)}{\uc(\vec r)}$ (orange) and $\au{n}(\vec r)$ (blue) obtain from the simulations as function of $\gamma$  with $(\sigma; n)=(1,0.5)$. }
  \end{center}
\end{figure}

These two boundary estimates have been plotted in Fig \ref{fig:range_transition} and show good correspondence with the numerical simulations.
This shows also that the relevant quantity is in fact the extension of  $\frac{ \norm{\au{m}(\vec r)}}{\uc(\vec r)}$ (with $m=1$ or $n$) which does not necessarily follow the extension of the velocity field $\au{m}$. This is confirmed in Fig. \ref{fig:sigma_au} which represents the evolution of the standart deviation of  $\norm{\au{n}(\vec r)}$ (the normalized velocity in regime III) and the standart deviation of $\frac{\norm{\au{n}(\vec r)}}{\uc(\vec r)}$ as function of $\gamma$.
For low $\gamma$ values, both quantities decrease with $\gamma$. However $\frac{\norm{\au{n}(\vec r)}}{\uc(\vec r)}$ is non-monotonic and becomes increasing above a certain value of $\gamma$. This fact explains then the non-monotonic evolution observed in Fig. \ref{fig:range_transition}.b.

\subsection{Statistical properties of the flow field in the transient regime II}

The flow field in the transient regime has interesting statistical properties. 
One way to apprehend it, is to notice a similarity with the problem of percolation.
Indeed, as shown in Fig. \ref{fig:flow_example_homuc}, when $\meanu$ is increased, more and more regions satisfy the criterion $\norm{u} > \uc$.
Connected regions satisfying this criterion allows to define clusters, which become larger and more numerous as the pressure gradient increases.
If the field $\norm{\vec u}/\meanu = \norm{\au{}(\vec r)}$ were constant, the transition would occur at 
$$ \frac{\norm{\au{}(\vec r)}}{\uc(\vec r)} = \frac{1}{\meanu}, $$
which would correspond to a percolation problem. 
Similar behaviors are therefore expected, such as cluster fractality and the presence of criticality. 
However, it is important to recall that the problem is not strictly equivalent to percolation because the  field  $\frac{\vec u}{\meanu}$ is not constant in the transient regime.
In particular, the change in viscosity introduces correlations in the velocity field and thus changes   the shape of the clusters.
 
Here, the clusters are identified using a Hoshen-Kopelman algorithm \cite{hoshen76}. 
Their shape are then characterized by their total size $S$ and the two dimensions of the bounding rectangle that contains it: $L$ along the flow direction and $W$ transversely to it.

\subsubsection{Size distribution} 

\begin{figure}[ht] 
\begin{center}
         \includegraphics[width=0.9\hsize]{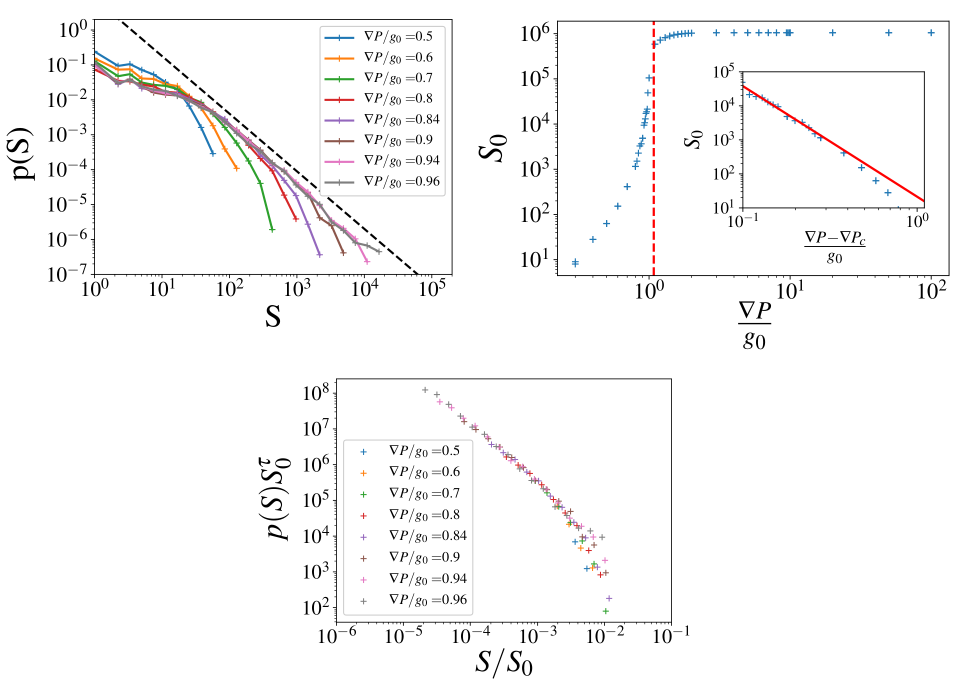}
  \caption{\label{fig:PS_Scut}  Left: Probability distribution function of the cluster size for different applied pressure drops using the parameters $(n,\gamma,\sigma_e)=(1/2,1/2,1)$. Dashed black line represents the fitted power-law exponent. Right : Size of the largest cluster $S_0$ versus the pressure gradient. The dotted vertical line represents $\nabla P_c$. Inset: $S_0$ as a function of $\langle \nabla P \rangle - \nabla P_c$ on a logarithmic scale, the line represents the fitted  power law. Bottom: Size probability distribution function for different pressure gradients normalized according to eqs. (\ref{eq:S0_GPC}) and (\ref{eq:PS}).
  } 
  \end{center}
\end{figure}    

Fig \ref{fig:PS_Scut}.a displays the size distribution $P(S)$ for different mean pressure gradient and for the parameters $n=1/2$ and $\sigma=1$. 
For any applied pressure, the distribution follows a decaying power law over a wide range of sizes.
However, a large-scale cutoff, $S_0$, is found that varies with the pressure gradient.
The plotting of the variation of $S_0$ as a function of pressure (Fig. \ref{fig:PS_Scut}.b) shows that $S_0$ diverges at a certain value of the average pressure gradient $\langle \nabla P \rangle = \nabla P_c$ according to a power law:
\begin{equation}
\label{eq:S0_GPC}
 S_0 \propto |\nabla P - \nabla P_c|^{-\nu_S}.
\end{equation}

Combining both observations, it follows the scaling law for the size distribution:
\begin{equation}
    p(S) \propto S^{-\tau_S} f(\frac{S}{S_0}),
    \label{eq:PS}
\end{equation}
which is confirmed by the good overlap of the distributions rescaled according to eqs. (\ref{eq:S0_GPC}) and (\ref{eq:PS}) plotted in Fig. \ref{fig:PS_Scut}.c.

This scaling law is thus similar to the one found in other problems with a critical transition such as  percolation \citep{stauffer91}, avalanches (e.g \cite{amaral95, santucci11}) or yield-stress fluid in porous media \cite{chevalier17,kostenko19}.
The sizes are distributed on many scales up to a size limit. And this size limit diverges as the control parameter, $\langle \nabla P \rangle$, approaches a critical value $\langle \nabla P \rangle = \nabla P_c$. The system then exhibits an infinitely broad range of scales (\emph{e.g.} fractal). This scaling law is characterized by the two exponents $\tau_S$ and $\nu_S$.

A similar scaling law could also be observed for the length $L$ of the clusters:
 \begin{eqnarray}
    p(S) \propto S^{-\tau_{L}} f(\frac{L}{L_0}) & \;\;\; \rm{with} \;\;\; & L_0 \propto |\langle \nabla P \rangle - \nabla P_c|^{-\nu_L}, 
    \label{eq:PL}
\end{eqnarray}
allowing to also identify the exponents $\tau_L$ and $\nu_L$.

For each set of parameters $(\sigma_f, \gamma, n)$ this scaling law is observed leading to the determination of the exponents  $\tau_S$, $\tau_L$, $\nu_S$ and $\nu_L$.
Tables \ref{tab:exp_sigma}-\ref{tab:exp_gamma} report these exponents according to the different sets of parameters.
The most remarkable result is the fact that $\tau_S$ seems to be indeed independent of the parameters. This is a characteristic found in many critical systems, where some exponents are independent of the details of the disorder distribution.  Such behavior is often referred to as "universal" as, for example, the exponents $\tau$ and $\nu$ in the percolation or avalanches of an elastic line in a random medium \citep{barabasi95}.
The results obtained here seem to suggest a universal behavior for the exponent $\tau_S$.
The observed value $\tau_S = 1.65 \pm 0.05$ is however very different from the standard percolation problem ($\tau_{perc} = 2.05$), which indicates that it would be of a different universality class.

The trend is less clear with the exponent $\nu_S$, which seems to vary with the rheological index $n$ and also the heterogeneities $\sigma_f$.
However, it should be noted that the determination of the exponent $\nu_S$ is generally more prone to errors because it requires the determination of $\nabla P_c$, which is also subject to uncertainties.
The error can be estimated at about $10\%$. It is then difficult to conclude on the universality of this exponent.

\subsubsection {Cluster's shape}

\begin{figure}[ht]
            \includegraphics[width=0.9\hsize]{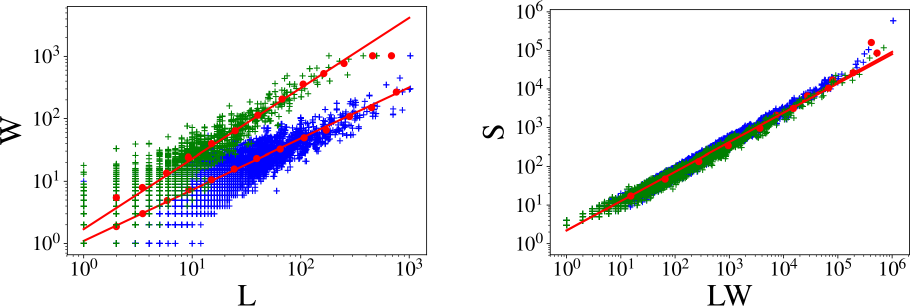} 
	\caption{\label{fig:shape}
	Left: Width ($W$) of clusters according to their length ($L$). Right: surface of the cluster $S$ according to the bounding size $WL$.
	Blue: $(n,\gamma,\sigma_f) = (0.5,0.75,1)$ and green:  $(n,\gamma,\sigma_f) = (0.5,0.25,1)$.
	The red circles correspond respectively to the average $W$ for a given $L$ (left) and the average $S$ for a given $LW$ (right). The red lines correspond to the power-law fit.
	}
    \end{figure}

\begin{figure}[ht]
    \includegraphics[width=0.9\hsize]{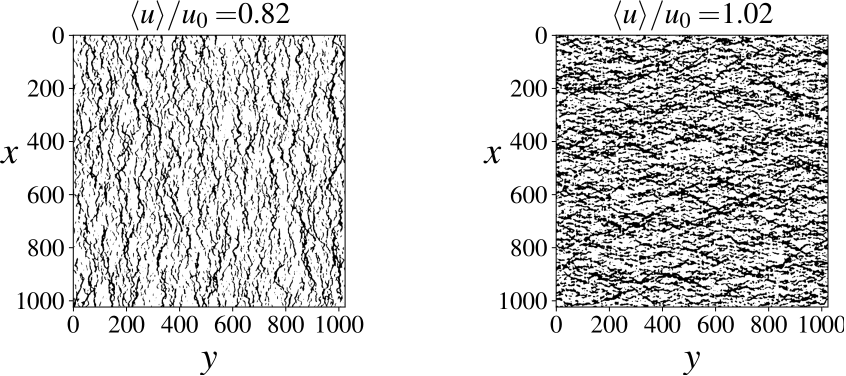} 
	\caption{\label{fig:shape_example} 
	Regions above the threshold $u>\uc$ (in black) for two different sets of parameters and close to the critical point. The mean flow direction is from top to bottom. Left: $(n,\gamma,\sigma_f) = (0.5,0.25,1)$ and right:  $(n,\gamma,\sigma_f) = (0.5,0.75,1)$. The $\gamma$ parameter drastically affects the orientation of the correlation from the direction of the flow to perpendicular to it.	
	}
    \end{figure}

The shape of the clusters can be characterized by the aspect ratio $L/W$.
Fig. \ref{fig:shape}.a shows the width $W$ of each cluster as a function of their length $L$ for two different sets of parameters. A power-law type relationship is then observed:
\begin{equation}
    \langle W \rangle_L \propto L^\zeta,
\end{equation}
which is a characteristic of the self-affine fractal structure typically found in anisotropic critical systems (e.g. avalanches, directed percolation, front propagations, etc.).
 It characterizes the fact that, although many different cluster sizes are present, the aspect ratio is not the same at each scale.
If $\zeta<1$, larger clusters are more elongated than the smaller ones as represented in Fig. \ref{fig:shape_example}.a.
For $\zeta>1$ bigger clusters are more elongated in the direction transverse to the flow (Fig. \ref{fig:shape_example}.b).

Because clusters are not compact, viz. they may contain holes, another interesting quantity  to analyze is the surface area ($S$) as a function of the enclosing box size $WL$, as shown in Fig. \ref{fig:shape}.b.
Here, again the relationship observed is a power-law: $$ S \propto (LW)^{\beta},$$ which is also a characteristic of a fractal structure.

The measured exponents for different sets of parameters, $\sigma$, $n$ and $\gamma$, are displayed in  table \ref{tab:exp_sigma}-\ref{tab:exp_gamma}.
The exponent of size $\beta$ seems to be almost constant, within the error bar, for any parameter value: $\beta \simeq 0.77 \pm 0.05$. This exponent seems to be universal.
More surprising is the evolution of $\zeta$ which takes only two values: either $\zeta \simeq 0.85 \pm 0.05$ or $\zeta \simeq 1.15 \pm 0.05$ depending on the parameters.
The two cases shown in  Fig. \ref{fig:shape_example} are in fact the only two observable exponents. 
In this figure, the two cases appear to be very similar, as if they were rotated by $90^\circ$.
In fact, such rotational symmetry can be proven when the rheological parameter $n$ is modified.
As detailed in the  \ref{sec:rotation}, rotation by $90^\circ$ of the velocity field and pressure field gradient is equivalent to solving the flow with inverse rheology (\emph{i.e} $n \rightarrow 1/n$).

We can therefore expect that, by reversing $n\rightarrow 1/n$, the shape of the clusters remains the same but rotated by $90^\circ$. This corresponds to the inversion of the roles of $W$ and $L$.
It thus leads to the relation: $$\zeta(n) = \frac{1}{\zeta(1/n)},$$ which seems to be satisfied by the numerical observations. The value $\zeta(n=1) \simeq 1$  appears therefore as a marginal value.
It remains quite remarkable that this exponent is constant for any $n$ exponent of shear thinning (or shear thickening). 

More unexpected is the similar change when $\gamma$ is varied (Table \ref{tab:exp_gamma}).
The value $\zeta$ is constant until a certain value $\gamma \sim 0.6$, where it switches to its inverse value.
Some remarks can be made to interpret this switch.
First, this change in correlation confirms the fact that the spatial correlation of the velocity field, correlated in the direction of flow, does not necessarily follow the correlation of the field $\vec u(\vec r)/\uc(\vec r)$.
Moreover, the value of $\gamma$ at which the change occurs, seems to correspond to the non-monotonic change observed in Fig. \ref{fig:sigma_au}.a.
To understand this change, it is useful to consider the two extreme cases $\gamma=0$ and $\gamma=1$. 

Before, it should  be noted that in the non-linear Darcy's law eq. (\ref{eq:Carreau_Darcy_vec}), the criterion $\norm{\vec u} > \uc(\vec r)$ also corresponds to a criterion for local pressure gradient $\norm{\grad P(\vec r)} > \gc(\vec r)$, with $\uc = A \kappa^{\gamma}$ and $\gc = \mu A \kappa^{\gamma -1}$.

For $\gamma =0$, $\uc$ is a constant: the cluster, defined by $\norm{\vec u} > \uc$, are expected to follow the correlation of the velocity field. The cluster should then be more elongated in the streamwise direction.
For $\gamma=1$, the situation is different because $\uc(\vec r)$ is now a random variable, but $\gc$ is  a constant field.
The clusters, equivalently corresponding to $\norm{\grad P(\vec r)} > \gc$, should then have a similar direction of correlation as $\grad P$.
In general, the pressure gradient tends to be correlated in the direction transverse to the flow\footnote{This can be seen in Appendix from the 90 degree rotation symmetry or the expansion perturbation}. 
The clusters are then expected to have a correlation direction transverse to the flow.
Depending on the $\gamma$ values, the shape of the cluster is then the result of a balance between these two opposite effects. 

If the cluster elongation changes with $\gamma$, it is still quite surprising that the correlation exponent $\zeta$ is constant and only switches from one value to another. This suggests that the exponent is also universal but the principal  direction of correlation is determined by the value of $\gamma$.

    \begin{table}
      \begin{center}
        \def~{\hphantom{0}}
            \begin{tabular}{cccc}
                $\sigma_f$ & $0.5$ & $1.0$ & $2.0$ \\
                \hline
                $\tau_S$ & 1.65 & 1.65 & 1.67\\
                $\nu_S$ & 3.0 & 3.25 & 3.2 \\
                $\nu_L$ & 1.8 & 1.8 & 1.75 \\
                $\zeta$ & 0.85 & 0.85 & 0.85\\
                $\beta$ & 0.77 & 0.77 & 0.77\\
            \end{tabular}
    \caption{\label{tab:exp_sigma}
    Measured exponents for different amplitude of heterogeneity $\sigma_f$ with 
    $(n,\gamma) = (0.5,0.5)$.}
    \end{center}
   \end{table}

       \begin{table}
      \begin{center}
\def~{\hphantom{0}}
    \begin{tabular}{cccccccc}
    $n$ & $1/3 $ & $0.5 $ & $2/3 $ & $1 $ & $2 $ & $3 $\\
    \hline
    $\tau_S$ & 1.65 & 1.65 & 1.65 & 1.65 & 1.65 & 1.65\\
    $\nu_S$ & 3.3 & 3.25 & 3.5 & 3.6 & 3.7 & 3.9\\
    $\nu_L$ & 1.95 & 1.8 & 1.8 & 1.8 & 1.9 & 1.9\\
    $\zeta$ & 0.83 & 0.85 & 0.85 & 0.95 & 1.15 & 1.17\\
    $\beta$ & 0.77 & 0.77 & 0.8 & 0.8 & 0.78 & 0.78\\
    
  \end{tabular}
  \caption{\label{tab:exp_n} 
  Measured exponents for different amplitude of rheological index $n$ with 
    $(\sigma_f,\gamma) = (1,0.5)$.
    }
    \end{center}
   \end{table}
   
          \begin{table}
      \begin{center}
\def~{\hphantom{0}}
    \begin{tabular}{ccccccc}
    $\gamma$ & $0 $ & $1/4 $ & $1/2 $ & $2/3$ & $3/4$ & $1 $\\
    \hline
   $\tau_S$ & 1.65 & 1.65 & 1.65 & 1.65 & 1.65 & 1.65\\
   $\nu_S$ & 3.35 & 3.3 & 3.25 & 3.15 & 3.3 & 3.4\\
   $\nu_L$ & 2.1 & 2.1 & 1.8 & 1.9 & 1.9 & 1.9\\
   $\zeta$ & 0.85 & 0.85 & 0.85 & 1.15 & 1.15 & 1.13\\
    $\beta$ & 0.77 & 0.77 & 0.77 & 0.77 &0.77 & 0.7\\
     \end{tabular}
  \caption{\label{tab:exp_gamma} Measured exponents for different $\gamma$ with 
    $(n,\sigma_f) = (0.5,1)$. }
    \end{center}
   \end{table}

\section{Conclusion \label{sec:conclusion}}

In this paper, we have studied the flow in macroscopic heterogeneous porous media with nonlinear rheology exhibiting a change in behavior such as truncated rheology. The influence of heterogeneities in permeability and velocity threshold fields was considered.

By varying the mean pressure gradient, three flow regimes are observed.
At low pressure gradient, the whole system is in the constant viscosity regime. The total flow rate then increases linearly with the mean pressure gradient.
At a high pressure gradient, when the entire system is in the nonlinear viscosity regime, the total flow rate increases non-linearly with the same exponent.
Transiently, the different regions of the medium change their viscous regime inhomogeneously with the mean pressure gradient, which induces a progressive change of the mean flow.

The first and last regime are relatively easier to analyse since they obey to the equation of a power-law fluid. This allow for instance a good prediction using a perturbation expansion approach.

The intermediate regime is more complex. Qualitatively, the pressure range of this regime is expected to be related to the width of the velocity distribution in regime I and III and thus to the amplitude of the permeability heterogeneities.
For a wider velocity distribution, the transition starts at a lower average velocity and ends at a higher velocity. 
Such behavior could be used to quantify the heterogeneities of a field for example.
While this trend was indeed observed, the results showed that it is more complicated to make quantitative predictions.
Indeed, a very important aspect of the problem is the relationship between velocity, permeability and local threshold.
In this paper, a power-law dependency between the permeability and the threshold has been assumed. The exponent $\gamma$ modifies drastically the correlation of the clusters.

The reason behind this observation is that different mechanisms are at work.
First, regions with high permeability have a higher velocity field but also a higher velocity threshold. If the velocity threshold varies weakly with permeability, the correlation follows that of the velocity field, in the direction of flow.
Conversely, if the velocity threshold varies strongly with permeability (i.e. high $\gamma$), the transition may not correspond to the highest velocities regions.
But at the same time, a similar reasoning can be made by considering the pressure field gradient: low permeability regions are more likely to have a higher pressure gradient but also a higher pressure threshold. The correlations could therefore be dominated by the pressure gradient field, transverse to the flow direction. The competition between these two effects is thus balanced by the $\gamma$ exponent. 
 
The relationship between the permeability and the two thresholds seems then to be very important. Here, a power law has been assumed but it is important to point out that in practice, the relationship is probably more complicated.

A remark can be made. In the literature, a very common approach is to model a porous medium or a fracture by a bundle of tubes or parallel layers \cite{federico98, chen05, nash16,felisa18}. This facilitates the analysis of the influence of heterogeneity and leads to a good qualitative understanding of the problem.
In these models, the flow field is by construction infinitely correlated in the streamwise direction, whereas the pressure gradient is uncorrelated in the crossflow direction. These models are thus expected to be unable to capture behaviors associated with the change in correlation. These models should therefore be taken with caution when applied to a 2D or 3D medium.

The statistical properties of the flow field in the transient regime also appeared very rich.
Indeed, in this regime, regions above their threshold define clusters that exhibit fractal and critical properties.
An important result is that some of the exponents (size distribution, shape exponents) do not vary with the parameters and the amplitude of the heterogeneity, which tends to suggest the presence of universal behaviors.
Another interesting feature is that the shape exponent $\zeta$ is constant but only switches to its inverse depending on the value of $\gamma$ and $n$.

This statistical feature is thus very reminiscent to related problems such as percolation \cite{stauffer91} and yield stress fluid in porous media \cite{kostenko19}.
The flow structure of yield stress fluid is, however, quite different because below the threshold there is no flow. Regions above the threshold are then necessarily channel paths connecting the inlet to the outlet. The exponents are thus expected to be  different. 
For yield stress fluid, it was found $\tau_S \simeq 1.15$ and $\zeta \simeq 0.75$. The size distribution exponent $\tau_S$ is different while the roughness exponent is similar to the present case.
As discussed previously, a slightly closer problem could be the percolation (directed or not), but the correlation of the velocity field evolves with the applied gradient of pressure.
The observed exponent are indeed different: $\tau_S \simeq 2.1$ for percolation and $\tau_S \simeq 1.26$ for directed percolation. The present case seems to be intermediate and if the exponents are ``universal'' they fall in a different universality class.

There are many interesting directions to pursue  this work. One important question is how to generalize it to the 3D permeability field.
Although rotational symmetry is no longer applicable in 3D, there remains the important fact that the velocity field is correlated along the stream direction while the pressure gradient is correlated in both transverse directions. 
Thus, a change in correlation should still be expected depending on $\gamma$. Critical behaviors should also probably be observed but with different exponents.
Another interesting study would be to investigate the problem of species transport in such a system.
Indeed, the dispersion of a tracer depends on the heterogeneity of the velocity field and its correlation. It is therefore expected to observe a change in behavior due to the change in rheology.
Furthermore, since the molecular diffusion coefficient is generally related to the viscosity of the fluid, it is expected to be different when the viscosity is below or above the threshold.
Finally, another direction of investigation could be other porous media problems with similar behaviors.
For example, the problems of two-phase flow \citep{tallakstad09a, yiotis13,sinha17,yiotis19}, emulsion driven in a porous medium \citep{leblay20} or  erosion of a granular bed \citep{aussillous16}, present a similar critical behavior with the appearance of preferential flow paths as a function of flow rate.
These problems have in common the property that at a certain critical velocity, the local flow conditions are drastically modified, because the bubbles are mobilized or because the grains rearrange themselves.
It would therefore be very interesting to study the similarities and differences between these problems.

\subsection*{Acknowledgements. ---}
This work is supported by "Investissement d'Avenir" LabEx PALM (ANR-10-LABX-0039-PALM). This work was partly supported by the Research Council, through its INTPART funding scheme, project number 309139.
I would like to thanks D. Salin,  R. Kostenko, A. Hansen and A. Rosso for fruitfull discussions.

\appendix

\section{Perturbation expansion for a power-law rheology \label{sec:expansion}}
In section \ref{sec:numerics}, Fig. \ref{fig:flowrate_example}, we have seen that in regimes I and III, the moments of the velocity distribution (mean, standard deviation, etc.) are constant.
In these regimes, the flow is governed by Darcy's law for a power law fluid in a heterogeneous medium. In this case, the mean and standard deviation can be determined using a perturbative approach. 
 Following the work of Gelhar and Axness \cite{gelhar83} for Newtonian fluids, the principle is to expand the solution around the mean value and assuming sufficiently small perturbations (\emph{i.e} $\sigma$ small).

The flow field is assumed to be solution of a power-law rheology in heterogenous porous media in the form:
\begin{equation}
\label{eq:power-law_with_C}
- \grad P = c(\vec r) \norm{\vec u}^{n-1} \vec u,
\end{equation}
with the free divergence:
\begin{equation}
 \nabla.\vec u =0.
\end{equation}

The field $c(\vec r)$ is assumed to be distributed according to a log-normal.
The principle is to decompose each field $\vec u$, $P$, and $c(\vec r)$ into a mean part and the spatially fluctuating part 

\newcommand{\cprime}{\delta c}
\newcommand{\gprime}{\delta g}

\begin{eqnarray}
 \vec u(\vec r) &=&  U \vec e_x +  \vecuprime (\vec r), \\
 \grad P(\vec r) &=& G \vec e_x + \grad \pprime (\vec r), \\
 c(\vec r) &=& C_0 + \delta C(\vec r).
\end{eqnarray}
 To simplify the notations, we introduce $G=\langle \nabla P \rangle$ and $U=\langle u_x \rangle$. The mean flow is assumed to be along the $\vec e_x$ axis.
 
It is more convenient to use the $g$ field defined by:
$$ c(\vec r) = \exp{(\bar g + \gprime)}= C_0 \exp(\gprime).$$

Expanding up to the second order, one have:
\begin{eqnarray*}
 c &=& C_0 ( 1 + \gprime + \frac{1}{2} \gprime^2) \\
 \norm{\vec u}^{n-1} &=& \norm{ U \vec e_x + \vecuprime}^{n-1} \\
 &=& U^{n-1} + (n-1) U^{n-2} \uprime_x + \frac{n-1}{2} U^{n-3} (\vecuprime.\vecuprime) \\
 & &        + \frac{(n-1)(n-3)}{2} U^{n-3} \uprime_x^2.
\end{eqnarray*}

Expanding and taking the spatial average $\langle . \rangle$ of eq. (\ref{eq:power-law_with_C}), 
all the first order terms vanish by definition. It yields:
\begin{eqnarray*}
 - G \vec e_x &=& C_0 U^n \vec e_x + \frac{1}{2} C_0 \langle \gprime^2 \rangle U^n \vec e_x + C_0  \frac{n-1}{2} U^{n-2}  \langle \vecuprime.\vecuprime \rangle \vec e_x      \\ & & +\frac{(n-1)(n-3)}{2} U^{n-2} \langle \uprime^2 \rangle+ C_0  U^{n-1} \langle \gprime \; \vecuprime \rangle \\ & & + C_0 (n-1) U^{n-1} \langle \gprime  \; \uprime_x \rangle \; \vec e_x + C_0 (n-1) U^{n-2} \langle \uprime_x \; \vecuprime \rangle.
\end{eqnarray*}
Along the $\vec e_x$ axis, it follows:
\begin{equation}
\label{eq:mean_G_vs_correlation}
 -\frac{G}{C_0 U^n} = 1 + \frac{1}{2} \langle \gprime^2 \rangle + n \frac{1}{U} \langle \gprime \; \uprime_x \rangle + \frac{n(n-1)}{2}\frac{1}{U^2} \langle \uprime_x \uprime_x \rangle + \frac{n-1}{2} \frac{1}{U^2} \langle \uprime_y^2 \rangle.
\end{equation}

This expression relates the mean gradient $G=\langle \nabla P \rangle$ to the mean velocity of the flow $U=\langle u_x \rangle$, if we know the different cross-correlation terms of the spatially fluctuating fields, which are determined next.

The first order of eq. (\ref{eq:power-law_with_C}) gives:
\begin{equation}
 - \grad \pprime = C_0 U^{n} \gprime \; \vec e_x + C_0 (n-1) U^{n-1} \uprime_x \vec e_x +C_0 U^{n-1} \vecuprime.
\end{equation}
Thus,
\begin{equation}
-\frac{1}{C_0 U^{n-1}} \grad{\pprime} = U \gprime \vec e_x  + (n-1) \uprime_x \vec e_x + \vecuprime.
\end{equation}

Taking the curl leads to:
\begin{equation}
 0 = \partial_x(\uprime_y) - \partial_y ( U \gprime + n \uprime_x).
\end{equation}

It is now more convenient to write this equation in  Fourier space.
Defining $\hat{\vec u}(k_x,k_y)$ and $\hat g(k_x,k_y)$, respectively the Fourier transform of $\vecuprime$ and $\gprime$, gives:
\begin{equation}
i k_x \hat u_y - i k_y ( U \hat g + n \hat u_x) = 0.
\end{equation}
Using the free divergence in Fourier space:
\begin{equation}
k_x  \hat u_x + k_y \hat u_y =0,
\end{equation}
it follows the relationship between the fluctuation of $\vecuprime$ and $\gprime$, in Fourier space:
\begin{eqnarray}
 \label{eq:TFu_TFg}
 \hat u_x &=& - \frac{k_y^2}{n k_y^2 + k_x^2} \hat g \\
 \hat u_y &=&  \frac{k_y k_x}{n k_y^2 + k_x^2} \hat g.  \nonumber
\end{eqnarray}

From these expressions, it is possible to determine the different cross correlation terms using  Parseval's formula: $$ \iint  F \; H  \; dxdy = \iint  \hat F \; \hat  H^* \; dk_x dk_y, $$ for any field $F(\vec r)$ and $H(\vec r)$.

Thus,
\begin{eqnarray}
 I_1 &=& \frac{1}{U^2} \langle \uprime_x  \uprime_x \rangle = \iint \hat u_x \hat u_x^* \;dk_x dk_y = \iint \frac{k_y^4}{(n k_x^2 + k_y^2)^2} \; \hat g \hat g^*   \;dk_x dk_y \label{eq:I1}\\
 I_2 &=& \frac{1}{U} \langle \uprime_x  \gprime \rangle = \iint \hat u_x \hat g^* \;dk_x dk_y = -\iint \frac{k_y^2}{n k_x^2 + k_y^2} \; \hat g \hat g^*   \;dk_x dk_y  \label{eq:I2}\\
 I_3 &=& \frac{1}{U^2} \langle \uprime_y  \uprime_y \rangle = \iint \hat u_y \hat u_y^* \;dk_x dk_y = \iint \frac{k_y^2 \; k_x^2}{(n k_x^2 + k_y^2)^2} \; \hat g \hat g^*   \;dk_x dk_y.\label{eq:I3}
\end{eqnarray}
 
These equations are very general and should apply to any field distribution and correlation $c(\vec r)$, provided that the amplitude of the heterogeneities is small enough. 

Using now the particular distribution field $g(\vec r)$ with 
\begin{equation}
 \hat g \hat g^* = B^2 e^{-   2 \frac{k_x^2+k_y^2}{k_0^2}},
\end{equation}
where $B$ is a normalisation prefactor determined by $\sigma_g=\sqrt{\langle \gprime^2  \rangle}$.

The different correlation functions can be derived after some manipulation:
\begin{eqnarray}
 I_1(n) &=& \frac{\sigma_g^2}{2 \pi} \int_0^{2\pi} \frac{\sin^4 \theta}{(n \cos^2 \theta + \sin^2 \theta)^2} \; d\theta =\frac{1}{2} \sigma_g^2 \frac{1 + 2 \sqrt{n}}{ (1+ \sqrt{n})^2n^{3/2}  } \label{eq:I1}\\
 I_2(n) &=& -\frac{\sigma_g^2}{2 \pi}\int_0^{2\pi} \frac{\sin^2 \theta}{n \cos^2 \theta + \sin^2 \theta} \; d\theta = - \sigma_g^2 \frac{1}{\sqrt{n}+n}\\
 I_3(n) &=& \frac{\sigma_g^2}{2 \pi} \int_0^{2 \pi} \frac{\sin^2\theta \cos^2\theta}{(n \cos^2\theta+ \sin^2\theta)^2} \;d\theta = \sigma^2_g \frac{1}{2 (1+\sqrt{n})^2\sqrt{n}}. \label{eq:I3}
\end{eqnarray}

It is important to note here that these results are independent of the correlation length $\lambda$ and the shape of the correlation function. This is due to the fact that the correlation function is istropic, so that the integrals eqs. (\ref{eq:I1}-\ref{eq:I3} can be split into a function depending only on $\norm{k}$ multiplied by  another depending only on the angular coordinate $\theta$\footnote{The integral over $|k|$ is equal to $\sigma_g^2/(2\pi)$.}.

The first correlation term, Eq. (\ref{eq:I1}), is the standard deviation of the velocity field which quantifies the heterogeneity of the velocity field:
\begin{equation}
\frac{\sigma^2_{u_x}}{U^2} = \sigma_g^2   \frac{1 + 2 \sqrt{n}}{ 2 (1+ \sqrt{n})^2n^{3/2}  }.
\end{equation}

These expressions can be used in eq. (\ref{eq:mean_G_vs_correlation}) to determine the mean pressure gradient $G$ by imposing the mean flow rate $U$:

\begin{equation}
\label{eq:mean_G_sigma_f_A}
 - \frac{G}{C_0 U^n} = 1 +  \sigma_g^2 \frac{\sqrt{n} -1 }{2(\sqrt{n}+ n)}
\end{equation}

A similar procedure could be used to express the mean flow rate $U$ as function of the imposed gradient of pressure $G$.
The full calculation is left to the reader. The basic idea is to write the constitutive equation in the form $\vec u = - D \norm{\grad P}^{\alpha-1} \grad P$, with $\alpha=1/n$ and $D=C^{- \alpha}$. 
This results to\footnote{Another main difference is that, in the first order expansion, the divergent has to be taken instead of the curl to eliminate the velocity and then relating the fluctuation  in pressure as function of the fluctuation in $C$.}:
  
\begin{equation}
 - \frac{U}{D_0 G^\alpha} = 1 +  \sigma_g^2  \alpha^2 \frac{\sqrt{\alpha} -1 }{2(\sqrt{\alpha}+ \alpha)}
\end{equation}
with $D_0 = C_0^{-\alpha}$.

Surprisingly, this expression is very similar to eq. (\ref{eq:mean_G_sigma_f_A}).
This similarity in fact  originates from a symmetry property of 2D flow fields where the role of pressure and velocity can be switched. This will be demonstrated and discussed below.

\section{Symmetry by a 90 degree rotation \label{sec:rotation}}

This argument originates from Matheron in a two-dimensional flow field for the Newtonian Darcy law  \cite{matheron67}. It can however be generalized to non-Newtonian fluids. This symmetry is also applicable to the 2D pore network model \citep{straley77,talon20}.
As previously, a generic non-linear Darcy equation is assumed:

\begin{equation}
 \grad.\vec u =0,
\end{equation}
and
\begin{equation}
 \vec u = - \frac{ f( \norm{\grad P}) } {\norm{\grad P}} \grad P
 \end{equation}
 or
\begin{equation}
 \grad P  = - \frac{ g( \norm{\vec u}) } {\norm{\vec u}} \vec u.
 \end{equation}

The idea consists in rotating the two fields $\vec u$ and $\grad P$ by $90^\circ$.
In a coordinate system $(x,y,z)$, where the flow takes place in the plane $(x,y)$, this rotation is performed by making the cross product with the vector $\vec e_z$:
\begin{equation}
 \vec e_z \times  \grad P =- \frac{ g( \norm{\vec u}) } {\norm{\vec u}} (\vec e_z \times \vec u).
\end{equation}
Defining the rotated fields $\vec q = \vec e_z \times \grad P$ and $\vec Z = \vec e_z \times \vec u$,
 it can be shown that $\grad . \vec q =0$ and $\grad \times \vec Z= \vec 0$.
 This means that $\vec q$ is a flux vector and $\vec Z$ derives from a potential field $\vec Z = \grad \Psi$.
The two new fields satisfy:
\begin{equation}
 \vec q = - g(\norm{\vec u}) \frac{ \grad \Psi }{\norm{\grad \Psi}}.
\end{equation}
Since $g(\norm{\vec u}) = \norm{\grad P} $, it follows  $\norm{\grad \Psi} = g^{-1}(\norm{ \vec q})$.
Yielding to:
\begin{equation}
 \grad \Psi = - g^{-1}(\norm{ \vec q } ) \frac{\vec q}{\norm{ \vec q}}.
\end{equation}
As a result,  the rotated fields $\vec q$ and $\grad \Psi$ satisfy a non-Newtonian Darcy's equation  but with a rheology inverse to the original one. In particular, solving a shear-thinning fluid in one direction is then equivalent to solving a shear thickening in the other direction.

Considering the truncated model:
\begin{equation}
\left\{
\begin{array}{lll}
\grad P = - \frac{\visc}{\kappa} \vec u & \; \; \; {\rm if}  \; \; \;& \norm{\vec u} < \uc \\
\grad P  = - \frac{\visc}{\kappa} \left[ \frac{ \norm{\vec u}}{\uc} \right]^{n-1} \vec{u} & \; \; \; {\rm if}  \; \; \;& \norm{\vec u} > \uc \\
\end{array}
\right.,
\end{equation}
 is thus equivalent to:
\begin{equation}
\label{eq:Carreau_Darcy_vec2}
\left\{
\begin{array}{lll}
 \grad \Psi = - \frac{\kappa}{\mu} \vec q  & \; \; \; {\rm if}  \; \; \;& \norm{\vec q} < q_c \\
\grad \Psi  = - \frac{\kappa}{\mu} \left[ \frac{ \norm{\vec q}}{q_c} \right]^{1/n-1} \vec{q} & \; \; \; {\rm if}  \; \; \;& \norm{\vec q} > q_c \\
\end{array}
\right., 
\end{equation}
with $q_c = \frac{\mu \; \uc}{\kappa}$.
Using the relation $\uc =A \kappa^{\gamma}$, the parameters have thus changed to:
\begin{eqnarray*}
 \tilde n &\rightarrow& 1/n \\
 \tilde \mu &\rightarrow& 1/\mu \\
 \tilde \kappa &\rightarrow& 1/\kappa \\
 \tilde \gamma &\rightarrow& -\gamma +1 \\
 \tilde A &\rightarrow& \mu A.
\end{eqnarray*}

It is remarkable that for the most natural value $\gamma=1/2$ in porous media, this coefficient is invariant with this transformation.
It should also be noted that the inverse of a lognormal distribution remains  lognormal with $\tilde f_0 \rightarrow - f_0$ and the same $\sigma_f$. It follow that the study can be limited to shear thinning fluids ($n<1$) without loss of generality.

\section{Numerical method\label{sec:numerical_method}}

\subsection*{Augmented Lagrangian method}
This Augmented Lagragian method has been introduced to solve non-Newtonian Stokes equation and has been used by many authors (see for instance \cite{glowinski89,roquet03}).
In this paper the method was adapted to solve the non-linear Darcy's equation:

\begin{eqnarray}
\label{eq:darcy_formal2}
\grad P &=& - g( \norm{\vec u} ) \frac{\vec u}{\norm{u}} , \\ 
\grad.\vec u &=& 0 \nonumber.
\end{eqnarray}
As for boundary conditions, pressure is imposed at the inlet and outlet, $P_{in}$ and $P_{out}$ respectively. Periodic conditions are assumed at the lateral sides (in the $y-$direction).
As described above, the solution of such system of equations is equivalent to finding, among all admissible velocity fields the minimum of the functional:
\begin{equation}
 \Phi[\vec u] = \int G(\norm{\vec u}) dxdy - P_{in} \int_{in} \vec u_x {dy} +  P_{out} \int_{out} u_x {dy}    ,
\end{equation}
with
$$G(\norm{\vec u}) = \int_0^{\norm{\vec u}} g(v) dv.$$  

The main idea of the augmented Lagrangian method is to introduce a secondary field $\vec v$ in order to decouple the nonlinear problem $G(\norm{{\vec u}})$ from the flow equation.
The equality $\vec u = \vec v$ is then guaranteed by the introduction of a Lagragian vector field $\vec \xi$.
An extra term $\frac{\alpha}{2}(\vec u - \vec v)^2$ is also added to enhance the convergence, where $\alpha$ is a small parameter.
Another Lagragian field $\chi$ is introduced to impose the free divergence, the problem can thus be recast into a saddle point determination:
\begin{equation}
 \min_{\vec u, \vec v} \max_{\chi,\vec \xi} \Psi[\vec u,\vec v,\vec \xi, \chi],
 \end{equation}
 with
\begin{eqnarray} \Psi[\vec u,\vec v,\vec \xi, \chi] &=& 
  \int \left[ G(\norm{\vec u}) + \vec \xi.(\vec v - \vec u)  + \frac{\alpha}{2}(\vec u - \vec v)^2 - \chi \vec \nabla. \vec v
\right] dxdy \\ & &
- P_{in} \int_{in} v_x {dy} +  P_{out} \int_{out} v_x {dy}.  \end{eqnarray}

If we now differentiate this functional, we obtain:
\begin{equation}
\label{eq:dPsi_dv}
\forall \delta \vec v,\;\;\;  \frac{\delta \Psi}{\delta \vec v}.\delta \vec v = 
\int \left[ \vec \xi + \alpha \vec v - \alpha \vec u \right].\delta \vec v- \int \chi \vec \nabla \delta \vec v \; dxdy -  P_{in} \int_{in}  \delta v_x {dy} +  P_{out} \int_{out} \delta v_x {dy},
\end{equation}
\begin{equation}\label{eq:dPsi_dlambda}
\forall \delta  \chi,\;\;\;  \frac{\delta \Psi}{\delta \chi} \delta \chi = - \int \delta \chi \; \vec \nabla \vec v \; dxdy
\end{equation}
\begin{equation}\label{eq:dPsi_du}
\forall \delta  \vec u,\;\;\;  \frac{\delta \Psi}{ \delta \vec u} . \delta \vec u = \int \left[ g(\norm{\vec u}) \frac{\vec u}{\norm{\vec u}} - \vec \xi  + \alpha \vec u - \alpha \vec v  \right].\delta \vec u \; dxdy
\end{equation}
\begin{equation}\label{eq:dPsi_dmu}
\forall \delta  \vec \xi,\;\;\;  \frac{\delta \Psi}{ \delta \vec \xi} . \delta \vec \xi = \int (\vec v - \vec u).\delta \vec \xi dxdy.
\end{equation}

This set of equations is quite cumbersome. However, the main advantage of this approach lies in the fact that, for a given $\vec u$ and $\vec \xi$,  finding the saddle point for $\vec u$ and $\chi$ eqs. (\ref{eq:dPsi_dv}-\ref{eq:dPsi_dlambda}), is equivalent to solving:
\begin{eqnarray}
\label{AL_Darcy_step}
 && \vec v = -\frac{1}{\alpha} (\grad \chi  + \vec \xi) + \vec u \\
 && \vec \nabla . \vec v =0,
\end{eqnarray}
with the boundary conditions $\chi= P_{in}$ and $\chi = P_{out}$.
This is a classical linear Darcy equation with a source term, which can therefore be solved using classical methods. Here, a second order finite difference method has been used.
It is important to note that in this equation the permeability is homogeneous and constant, which allows a very fast solution at each step.

For the given field $\vec v$, $\chi$ and $\vec \xi$, the minimization of $\Psi$ with respect to $\vec u$, is equivalent to solve:
    \begin{equation}
    \label{eq:AL_implicit}
     (\frac{g(\norm{\vec u})}{\norm{\vec u} } + \alpha ) \vec u = \vec \xi + \alpha \vec v,
    \end{equation}
 which represents an implicit problem. This can be solve numerically or analytically.
 With our particular function $g(\norm{u})$, an analytical solution can be found for some exponents $n= 1/3, 1/2, 2/3, 1, 3/2, 2 , 3$.
Solutions are given in a following section.
 
 After defining initial fields $\vec u_0$, $\vec v_0$, $\vec \xi_0$ and $\chi_0$, the algorithm is decomposed in the following step:
 
 \begin{itemize}
  \item[1.] Solve $\vec v_{n+1}$ and $\chi_{n+1}$ with the Darcy's equation:
    \begin{eqnarray}
        \label{eq:AL_Darcy_step}
        && \vec v_{n+1} = -\frac{1}{\alpha} (\grad \chi_{n+1}  + \vec \xi_{n}) + \vec u_{n} \\
        && \vec \nabla . \vec v_{n+1} =0,
    \end{eqnarray}
    Here, these equations are solved using a second order finite difference method.
  \item[2.] Determine $\vec u_{n+1}$, by solving 
    \begin{equation}
     (\frac{g(\norm{\vec u_{n+1}})}{\norm{\vec u_{n+1}} } + \alpha ) \vec u_{n+1} = \vec \xi_{n} + \alpha \vec v_{n+1},
    \end{equation}
    \item[3.] Advancing $\vec \xi$ toward the gradient eq. (\ref{eq:dPsi_dmu}):
        \begin{equation}
         \vec \xi_{n+1} = \vec \xi_{n} +(\vec v_{n+1} - \vec u_{n+1}) d\xi,
        \end{equation}
where $d\xi$ is a small parameter,  taken equal to $\alpha$ for simplicity.
 \end{itemize}

 \subsection*{Validation}

This section presents the validation of the numerical method, in particular, the influence of the mesh size.
In this problem, the main characteristic length is the field correlation length $\lambda$ which determines the amplitude of the velocity and pressure gradients.
The numerical resolution is then related to $N$, the number of mesh nodes per length $\lambda$.
In this paper, all the simulations were performed with $N=5$ and a total system size $1024\times1024$.

\begin{figure}
\centering
\includegraphics[width=0.9\hsize]{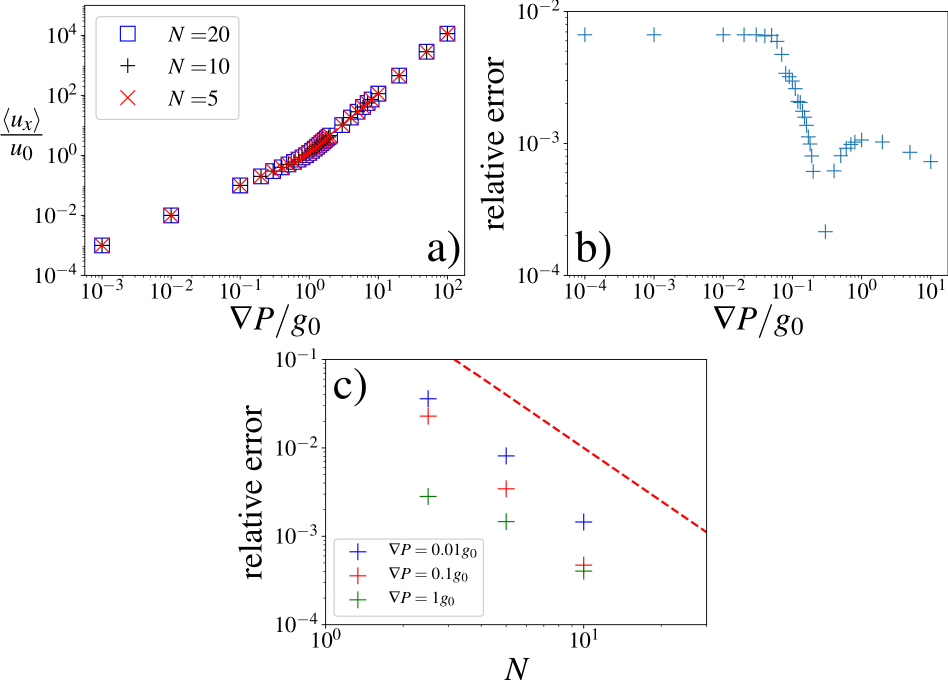} 
 \caption{\label{fig:valid}}  (a): Mean velocity as function of the mean gradient of pressure for different mesh resolutions. The parameters are $n=1/2$, $\gamma=1/2$ and $\sigma_f=1$.
 (b): relative error between $N=5$ and $N=20$ as function of $\nabla P$. (c): relative error, with the reference to $N=20$, as a function of $N$ and for different $\nabla P$.
The red dashed line corresponds  to the power law $N^{-2}$.
 \end{figure}

Fig. \ref{fig:valid}.a displays the mean flow rate as function of $\nabla P$, for different $N$. Although the difference between high and low resolution is not noticeable in this figure,  a relative error between  low  and higher resolution can be defined:
\begin{equation}
 {\rm relative\;  error}= \frac{\langle u_x \rangle^{N=5} - \langle u_x \rangle^{N=20}}{\langle u_x \rangle^{N=20}}
\end{equation}
Fig. \ref{fig:valid}.b represents the evolution of this error as a function of $\nabla P$.
The error depends on the flow regime and is, surprisingly,  more important in the linear regime than in the non-linear one. The maximum error, however, does never exceed $1\%$.
Figure \ref{fig:valid}.c shows the convergence of this error as function of the mesh resolution $N$ and different $\nabla P$. The convergence rate is then slightly faster than $N^{-2}$, which is in agreement with the second order finite difference scheme used.

\subsection*{Solutions of eq. (\ref{eq:AL_implicit}})
For the given fields $\vec v$ and $\vec \xi$, the alghorithm requires to find $\vec u$ satisfying:
    \begin{equation}
    \label{eq:AL_implicit_2}
     (\frac{g(\norm{\vec u})}{\norm{\vec u} } + \alpha ) \vec u = \vec \xi + \alpha \vec v \equiv \vec B,
    \end{equation}
with 
\begin{equation}
\left\{
\begin{array}{lll}
g(\norm{ \vec u}) =  \frac{\mu}{\kappa} \norm{ \vec u} & \; \; \; {\rm if}  \; \; \;& \norm{\vec u} < \uc \\
g(\norm{ \vec u}) =  \frac{\visc}{\kappa} \left[ \frac{ \norm{\vec u}^{n}}{\uc^{n-1}} \right] & \; \; \; {\rm if}  \; \; \;& \norm{\vec u} > \uc \\
\end{array}\right.
\end{equation}

It must be noted  that $\vec u$ and $\vec B$ are colinear and with the same orientation because the left term in eq. (\ref{eq:AL_implicit_2}) is positive. It is thus sufficient to determine the norm of $\norm{\vec u}=u$.
The equations then become:
\begin{equation}
\left\{
\begin{array}{lll}
  (\frac{\mu}{\kappa} + \alpha ) u = B & \; \; \; {\rm if}  \; \; \;& B < (\frac{\mu}{\kappa} + \alpha ) \uc \\
  A u^n + \alpha u = B & \; \; \; {\rm if}  \; \; \;&  B > (\frac{\mu}{\kappa} + \alpha ) \uc \\
\end{array}\right.,
\end{equation}
where $A = \frac{\visc}{\kappa \uc^{n-1}}$.

If the first equation is trivial, the second one has an analytical solution only for specific value of $n$. We give here the ones used in this work.

\begin{itemize}

 \item  $n=\frac{1}{3}$:
$$ u = -\frac{\sqrt[3]{\frac{2}{3}} A^3}{\Delta}+\frac{\Delta}{18^{1/3} \alpha ^3}+\frac{B}{\alpha }, $$
with   $$ \Delta =\left( \sqrt{3}\sqrt{4 A^9 \alpha ^9+27 A^6 \alpha ^{10}B^2}-9 A^3 \alpha ^5 B\right)^{1/3}.$$

 \item $n=\frac{1}{2}$: $$ u = \left( \frac{-A+\sqrt{A^2+4 \alpha B}}{2 A} {} \right)^2. $$

\item $n=\frac{2}{3}$:
   \begin{eqnarray*}
    \text{if} \;\;B < \frac{4 A^3 }{27 \alpha^2} , & &\\
    &&u = \frac{1}{\alpha^3} \left[ - (A^3-3\alpha^2 B) + 2 A^{3/2} \sqrt{A^3 - 6 \alpha^2 B} \cos{(\phi/3)}  \right],  \text{ with  }\\ &&\phi= \arg{\left[ -2 A^6 + 18 A^3 \alpha^2 B - 27 \alpha^4 B^2 + i 3 \sqrt{3} \alpha^3 \sqrt{4 A^3B^3 - 27 \alpha^2 B^4}\right]}.\\
    \text{if} \;\;B >\frac{4 A^3 }{27 \alpha^2} , & &\\
 &&  u = \frac{1}{3 \alpha^3} \left[ -(A^3 - 3 \alpha^2 B ) - 2^{1/3} (-A^6+ 6 A^3 \alpha^2 B) \frac{1}{\Delta} + \frac{\Delta}{2^{1/3}}  \right] ,  \text{ with  }\\
 &&    \Delta =  -2 A^6 + 18 A^3 \alpha^2 B - 27 \alpha^4 B^2 + 3 \sqrt{3} \alpha^3 \sqrt{-4 A^3B^3 + 27 \alpha^2 B^4 }.
    \end{eqnarray*}

 \item $n=2$:
 $$ u = \frac{- \alpha + \sqrt{\alpha^2 + 4 AB} }{2A} .$$
   
\item $n=3$:

$$ u = \frac{\Delta}{18^{1/3}
   A}-\frac{\sqrt[3]{\frac{2}{3}} \alpha }{\Delta}
,
$$ with
$$\Delta=\left( 9 A^2 B +\sqrt{3} \sqrt{27 A^4 B^2+4 A^3
   \alpha ^3} \right)^{1/3}.$$

\end{itemize}



\end{document}